\documentclass[12pt,preprint]{aastex}

\shorttitle{Spectroscopy of M81 GCs}
\shortauthors{Nantais and Huchra}

\begin{document}
\title{Spectroscopy of M81 Globular Clusters$^{1}$}

\author{Julie B. Nantais and John P. Huchra}
\affil{Harvard-Smithsonian Center for Astrophysics}
\affil{60 Garden Street, Cambridge, MA 02138, USA}

\footnotetext[1]{This study uses observations from the MMT Observatory, a joint facility of the Smithsonian Institution and the University of Arizona.}

\begin{abstract}
We obtained spectra of 74 globular clusters in M81.  These globular clusters had been identified as candidates in an HST ACS I-band survey.  68 of these 74 clusters lie within 7$\arcmin$ of the M81 nucleus.  62 of these clusters are newly spectroscopically confirmed, more than doubling the number of confirmed M81 GCs from 46 to 108.  We determined metallicities for our 74 observed clusters using an empirical calibration based on Milky Way globular clusters.  We combined our results with 34 M81 globular cluster velocities and 33 metallicities from the literature and analyzed the kinematics and metallicity of the M81 globular cluster system.  The mean of the total sample of 107 metallicities is $-1.06 \pm$ 0.07, higher than either M31 or the Milky Way.  We suspect this high mean metallicity is due to an overrepresentation of metal-rich clusters in our sample created by the spatial limits of the HST I-band survey.  The metallicity distribution shows marginal evidence for bimodality, with the mean metallicities of metal-rich and metal-poor GCs similar to those of M31 and the Milky Way.  The GC system as a whole, and the metal-poor GCs alone, show evidence of a radial metallicity gradient.  The M81 globular cluster system as a whole shows strong evidence of rotation, with V$_r$ (deprojected) = 108 $\pm$ 22 km s$^{-1}$ overall.  This result is likely biased toward high rotational velocity due to overrepresentation of metal-rich, inner clusters.  The rotation patterns among globular cluster subpopulations are roughly similar to those of the Milky Way: clusters at small projected radii and metal-rich clusters rotate strongly, while clusters at large projected radii and metal-poor clusters show weaker evidence of rotation.
\end{abstract}

\section{Introduction}
The study of globular clusters (GCs) is an excellent way to learn about the developmental history of nearby galaxies, for several reasons.  First, they provide a ``fossil record'' of a galaxy's most intense episodes of star formation.  Secondly, since GCs are close approximations of simple stellar populations (SSPs), it is relatively easy to estimate their ages and metallicities spectroscopically.  Also, having luminosities of $\sim 10^4-10^6 M_{\odot}$, they are much brighter than most individual stars and can thus be more easily studied photometrically and spectroscopically.   \citet{bro06} provide a comprehensive review of our understanding of GC systems (GCSs) and how they formed.  One common finding is that most elliptical and lenticular galaxies \citep{lar01,kun98,lee98} and early- and intermediate-type spiral galaxies \citep{zin85,hsv82} have bimodal metallicity distributions in their GCSs, comprised of an old metal-poor (MP) subpopulation and a slightly younger metal-rich (MR) subpopulation.  The mean metallicities of these MR and MP subpopulations \citep{str04} and the mean metallicity of the GCSs as a whole \citep{bro91} are found to vary systematically with the mass of the galaxy.

While GC metallicity distributions among galaxies of diverse masses and Hubble types show a systematic pattern, the kinematic properties of GCSs tend to be qualitatively different even among galaxies of similar mass and Hubble type.  For instance, while rotation is found among the MR and MP GC subpopulations of the Virgo giant ellipticals M87 \citep{cot01} and M60 \citep{hwa08}, the GCS of Fornax giant elliptical NGC 1399 does not appear to be rotating at all \citep{ric04}.  Such kinematic differences are also seen among disk galaxies.  The GCS of M31 shows strong rotation \citep{lee08,per02}, while in the Milky Way GCS there is only notable rotation among bulge GCs and very MP halo clusters \citep{har01}.  In contrast to both of these galaxies, M104, a nearby massive S0/a galaxy, shows no signs of net rotation of its GCS as a whole.  This result holds even in sub-populations of the M104 GCS divided according to metallicity or galactocentric distance \citep{bri07}.  Some late-type spiral galaxies in the Sculptor Filament show possible evidence of strongly rotating GCSs \citep{ols04}, but \citet{nan10a} did not find clear evidence of rotation in the GCS of the Sculptor galaxy NGC 300, and \citet{mor08} did not find notable rotation in a spectroscopic study of NGC 45.  These variable rotation patterns in the kinematics of GCSs suggest that similar galaxies may have had very different early formation histories that do not necessarily follow a neat pattern according to mass and/or Hubble type.

M81, an Sab spiral galaxy with a distance modulus of 27.7 \citep{fre01}, is an ideal target for furthering our understanding of spiral galaxy GCSs.  Three spectroscopic studies of M81's GCS have been performed thus far.  \citet{bro91} (hereafter BH91) observed eight GCs in M81 as part of a multi-galaxy study, with a variance-weighted mean metallicity of -1.46 (unweighted mean -1.08).  \citet{pbh95} (hereafter PBH) estimated the metallicities and radial velocities of 25 GCs from a sample of 82 candidates chosen from the \citet{pr95} (PR95) GC candidate catalog.  PBH found a mean metallicity of -1.48 $\pm$ 0.19, and no clear evidence of rotation, though this result can easily be explained by their having sampled mostly a MP halo population like that of the Milky Way. \citet{sch02}, hereafter SBKHP, found 16 more GCs, mostly MR and projected onto the M81 disk.  Combining all spectroscopically confirmed GCs from their own work, PBH, and BH91, SBKHP found clear evidence for GC rotation at intermediate distances of 4-8 kpc, but not in the inner 4 kpc or the outer regions.  The mean metallicity of the known GCs up to that time was -1.25 $\pm$ 0.13, similar to the Milky Way and M31.

In this study, we will extend the work of SBKHP and PBH by analyzing the spectra of 74 GCs, 62 of them never previously spectroscopically confirmed as clusters.  In \S 2 we describe our data.  In \S 3 we describe our methods of identifying objects.  In \S 4 we provide an analysis of metallicity, including a metallicity histogram and the metallicity of GCs as a function of distance.  In \S 5 we analyze the kinematics of the GCS, including the kinematics of subpopulations and a mass estimate for M81.  Finally, in \S 6 we summarize our findings.

\section{Data}
We obtained spectra of a total of 207 extended objects in M81 on the nights of 2006 May 3-5 and 2007 November 13 and 17 with Hectospec on the 6.5 m MMT on Mt. Hopkins in Arizona.  Hectospec has 300 fibers of $\sim$1{\arcsec} diameter.  The number of objects we could observe was limited by how many fibers could be placed within a small region; we could not observe more than $\sim$75 objects on and near the disk of M81 (within the range of the Nantais et al. 2009b HST I-band images) in a single pointing.  We used the 270 line mm$^{-1}$ grating, with spectral coverage from 3700-9150 {\AA}  and spectral resolution of about 5.1 \AA, similar to many other studies on extragalactic GCs.  The fibers covered a diameter of 1.5$\arcsec$ on the sky.

The objects observed on 2006 May 3 and 4 have total exposure times of 80 minutes (four 20-minute exposures), and the spectra from 2006 May 5 have total exposure times of 60 minutes (three 20-minute exposures).  Objects for these three exposures were selected from the \citet{nan10} catalog on the basis of magnitude, with bright objects preferred over faint objects.  The objects observed in November 2007 had 6 20-minute exposures on each night, but four of those exposures on November 17 were affected by clouds, so we combined 8 exposures, 6 from November 13 and 2 from November 17, to create the final spectra.  For the November 2007 observations, we gave extra priority to objects with an especially high likelihood of being GCs, based on both their \citet{nan10} g-r and r-I colors and their visual appearance.  We also prioritized observation of objects with low-quality or no prior spectra in either the May 2006 observation run or the existing literature.

Ideally, we would accomplish background subtraction by measuring the sky level using apertures on either side of the object.  This would account for both the temporal and spatial variation in the background.  However, we cannot do this with a fiber spectrograph.  Instead, we observed several sky offset frames each night, each taken about 5$\arcsec$ to the north, south, east, or west of the object exposures.  For the 2006 May 4 observations, only one sky offset exposure was used to sky-subtract these images, since the rest had sky fluxes either notably brighter or notably fainter than in the object exposures.  The spectra extracted using these unreliable sky exposures suffered from either over-subtracted or under-subtracted sky continuum that dominated the object's own spectral features.  On 2007 November 17, three of the sky exposures were lost to cloudy weather, so the total November 2007 sky offset spectrum was an average of 9 expsoures.  Table 1 lists the observation details for each night: the number of objects observed, the number of object pointings, the number of sky pointings, and the total integration time.

Each CCD image was flat-fielded, bias-corrected and wavelength-calibrated by the SPECROAD$^{2}$ automated pipeline.  This pipeline uses the ``hectospec'' package of IRAF tasks specially written at the Harvard-Smithsonian Center for Astrophysics (CfA) to facilitate the reduction of Hectospec data, plus tasks from other IRAF packages such as ``mscred'' and ``rvsao.''  All files used for bias subtraction, flat fielding, and wavelength calibration (lamp files) were taken on or as near as possible to the date of observation.  The automated pipeline followed to its full extent uses the nearest sky fiber from a given pointing in order to subtract the background, which does not account for the very high spatial variation in the background.  Therefore, in order to perform sky subtraction, we used object exposure spectra and sky offset exposure spectra produced by an earlier stage in the pipeline process in which sky subtraction and velocity determination had not yet been performed.  We averaged all individual exposures of both the object and the sky offset taken during acceptable weather conditions, and subtracted the average sky offset spectra from the average object spectra.  Flux calibration was then performed using a standard star spectrum observed on or as near as possible to the night of observation.  Any remaining cosmic rays and poorly subtracted telluric lines were removed individually from the final object spectra using the ``x'' command in the IRAF task specplot in the noao.onedspec package.  

\footnotetext[2]{See http://tdc-www.harvard.edu/instruments/hectospec/specroad.html for details on the SPECROAD pipeline.}

\section{Object Identification}

Objects were identified by a combination of radial velocity and visual inspection of their spectroscopic features.  Any object with a radial velocity greater than $\sim$ 400 km s$^{-1}$ and clear evidence of redshifted absorption or emission features was determined to be a background galaxy.  H{\small{II}} regions were recognizable by their emission lines near zero redshift.  GCs had a red spectrum and discernible old stellar population absorption features near zero redshift, and radial velocities less than 400 km s$^{-1}$.  Objects with no or very faint emission lines, a blue continuum, and absorption lines typical of a young (A-F) stellar population were labeled as open clusters.  The plausibility of a given object identification was confirmed via the HST I-band images.  No objects identified in our catalog as having GC-like morphology were spectroscopically determined to be young or open clusters.  Since our spectroscopic candidates were chosen based on size information in space-based imaging data, they are almost certain not to be stars.  Figure 1 shows the locations of different objects with respect to the M81 disk.  Figure 2 shows sample spectra of GCs, including a combined spectrum of all GCs.  We found a total of 74 GCs, 47 H{\small{II}} regions, 23 background galaxies, 7 open clusters, and 56 objects with too low signal-to-noise to determine their nature.  Of the 74 GCs, 12 have been previously observed: 8 by SBKHP, 3 by PBH, and one by BH91.  We have thus identified 62 newly-confirmed GCs.  Figure 3 shows the distribution of the radial velocities of all recognized objects - GCs, H{\small{II}} regions, and open clusters combined - within M81.  No recognized object that is not a background galaxy has a radial velocity greater than 400 km s$^{-1}$ or less than $-$400 km s$^{-1}$.  

One object, 1859 (Figure 4), showed both strong GC continuum and absorption features and notable H{\small{II}} emission.  Its HST image resembled a GC, with what may have been a small H{\small{II}} region some 2{\arcsec} away.  We therefore treat it as a GC, and remove the H{\small{II}} emission in order to calculate its radial velocity.  The GC absorption features yield a radial velocity of 114 $\pm$ 17 km s$^{-1}$, while the H{\small{II}} emission lines yield a radial velocity of 35 $\pm$ 65 km s$^{-1}$.

Table 2 shows the positions and radial velocities of our GCs and the 34 GCs from BH91, PBH, SBKHP that we did not re-observe.  The first few columns of Table 2 give the ID number in our catalog, in the PBH and SBKHP spectroscopic catalogs, and in the PR95 and \citet{cha01} (CFT01) photometric catalogs.  The next three columns give the RA in hours, the declination in degrees, and the projected galactocentric distance in arcminutes.  The final two columns in Table 2 give the radial velocity and its uncertainty in km s$^{-1}$.  Table 3 shows the positions, velocities, and object types of all HII regions, galaxies, and open clusters.  It is organized similarly to Table 2, but with an extra column after the ID number in our catalog listing the object type: galaxy (gx), HII region (hii), or open cluster (oc).

Our spectroscopic identifications of objects can be compared to the visual identifications in the I-band images in \citet{nan10}, as a check on the reliability of the visual classification scheme in the \citet{nan10} catalog.  Of the 107 GC candidates we observed, 80 had high enough signal-to-noise to confirm their nature; and of those 80 objects, 73 (91\%) were confirmed as GCs, 5 (6\%) were found to be H{\small{II}} regions, and 2 (3\%) were identified as galaxies.  Assuming these spectroscopic results are representative of GC candidates, the vast majority of our GC candidates in \citet{nan10} are true GCs.  We also observed 26 H{\small{II}} region/OB association candidates, 23 galaxy candidates, and 50 ``Other'' or unclassified objects.  All 26 H{\small{II}} region and OB association candidates were found to be either H{\small{II}} regions or open clusters; 24 (92\%) were confirmed as H{\small{II}} regions, and the remaining two objects were dominated by blue continuum and were classified as open clusters.  Of the 23 galaxy candidates, 17 had spectra with sufficient signal-to-noise to determine the object type, and all 17 were confirmed to be galaxies.  Of the 50 ``Other'' or unclassified objects, only 27 had sufficient signal-to-noise to be identified.  One of these spectra (3.7\% of the identified ``Others'') was confirmed as a GC, 17 (63\%) were confirmed as H{\small{II}} regions, 5 (18.5\%) were confirmed as open clusters, and 4 (14.8\%) were confirmed as galaxies.  These results, if typical of ``Other'' objects, suggests, as we found in \citet{nan10}, that the ``Other'' category is dominated by young objects such as H{\small{II}} regions, OB associations, and open clusters, with most of the remaining objects probably being galaxies.

\section{Metallicity Analysis}
Spectral indices and metallicity estimates were determined using the same methods and wavelength definitions as in \citet{nan10a}, which was based on the \citet{bro90} method but using indices calibrated on the \citet{sch05} Milky Way GC spectra smoothed to 5 {\AA} resolution. GC metallicites from \citet{har96}$^{3}$ were used for the calibration of index strength to metallicity.  Since we did not observe any Lick standard stars, we do not correct our fluxes to the Lick system.  We calibrated the full list of 25 indices used in \citet{nan10a}, and chose which ones to use on the basis of both the quality of the calibration and the scatter of the index in the M81 data.  We chose MgH, Mg2, Mgb, Fe5270, Fe5335, Fe5406, G4300, $\delta$, and CNR to determine a weighted average of the metallicity.  Tables 4 and 5 give the spectral indices we measured.  In Tables 4 and 5, the first column is the ID in our catalog, and the remaining columns give the values of the individual spectral indices named in the column header in magnitudes.  Below each row of data beginning with an ID number is a row labeled ``$\sigma$.''  Each ``$\sigma$'' row lists the 1$\sigma$ uncertainty in each index, in that index's column, for the object whose ID number is listed in the previous row.   Figures 5 and 6 show the spectral indices used for metallicity estimates as a function of the Mg2 index.  Table 6 gives the metallicites we calculated.  The table is arranged as three sets of three columns each.  The first column in each set of 3 (columns 1, 4, and 7) gives the object ID, the second column in each set (columns 2, 5, and 8) gives the metallicity, and the third column in each set (columns 3, 6, and 9) give the 1$\sigma$ uncertainty in the metallicity. Table 7 shows the linear fits to the Milky Way index-metallicity relations, with the index ID in the first column, the slope (a) and intercept (b) of the $[Fe/H] = a(index)+b$ in the next two columns, the correlation coefficient R$^2$ in the fourth column, and the R$_I$, $\sigma_m$, and $\sigma_s$ columns as defined in \citet{bro90} in the next three columns.  In Table 7 our $\sigma_s$ (uncertainty of repeat exposures) estimates are based on the largest dispersion of index values of the four individual exposures (in the same observing run) of the two brightest GCs: 2029 and 743.  The dispersions in the index were calculated for each object, and the largest of the two dispersions was adopted as $\sigma_s$.  Ideally, we would use several observations of the same high signal-to-noise object in two or more separate observing runs on two or more different nights, but we did not have such observations.  The index differences in repeat observations of objects on separate nights were dominated by the low signal-to-noise of these objects in one or both exposures; therefore, we concluded that the safest way to measure $\sigma_s$ was to use repeat observations of very high signal-to-noise objects on the same night.

\footnotetext[3]{We use the 2003 update of the catalog, located at http://physwww.mcmaster.ca/$\sim$harris/mwgc.dat}

Our own sample of 74 GCs has a mean metallicity of $-0.96 \pm 0.08$.  Beyond our own sample, there are also 34 GCs from SBKHP, PBH, and BH91 that we did not observe, 33 of which have spectroscopic metallicity estimates.  In \citet{nan10}, we raise the possibility that six of the PBH clusters and two of the SBKHP clusters may have been misidentified stars, but here we shall assume that all objects identified as clusters in the literature are indeed GCs.  For more complete analysis, however, we also perform metallicity, kinematics, and mass analysis excluding these 8 objects.  Combining our metallicities with those of the 33 clusters from the pre-existing literature, the mean metallicity of the M81 GCS is $<$[Fe/H]$>$ = $-1.06 \pm 0.07$.  If we exclude the 8 clusters flagged as possible stars from the total sample, we find a slightly lower mean metallicity, $<$[Fe/H]$>$ = $-1.09 \pm 0.07$.  The mean metallicity of the M81 sample, with or without the starlike objects, is comparable to the divide between the MR and the MP GCs in the metallicity histogram shown in Figure 7.  Figure 7 displays the metallicities of GCs in the M81, the Milky Way \citep{har96}, and M31 \citep{bar00} in 0.25 dex bins.  The divide in the M81 sample seems lower in metallicity than the Milky Way and M31 samples, and M81 appears to have somewhat of an excess of metallicities in the $-0.8$ to $-1.1$ range.  Rebinning the histogram with larger (0.2 dex) or smaller (0.3 dex) bins or shifting the histogram bin centers by 0.125 dex does not eliminate the apparent excess of objects at these metallicities or make the gap between high and low metallicities appear more similar to the Milky Way and M31.  

We performed a homoscedastic KMM test \citep{ash94} in order to statistically check for a bimodal metallicity distribution.  For this test, we excluded four extremely MR or MP metallicities with very high uncertainty values (low signal-to-noise).  The test provides marginal evidence for bimodality (P = 0.143) in the M81 globular cluster system, with subpopulation mean metallicities of -1.55 for the MP clusters and -0.61 for the MR clusters.  53 clusters are assigned to the MP group and 50 to the MR group, with an overall correct allocation rate of 87\%.  These groupings most likely do not represent the true proportions of MR and MP clusters in M81, due to our suspected lack of MP halo clusters.  The results are similar if the starlike clusters are excluded: P = 0.11 and subpopulation mean metallicities of -1.53 and -0.62.  The exclusion of the eight pointlike ``suspected stars'' from PBH and SBKHP also slightly reduces the bias in favor of the MR clusters, with 52 clusters being assigned to the MP group and 44 to the MR group, with an overall correct allocation rate of 88\%.

Figure 8 shows the GC metallicity vs. galaxy luminosity relations from \citet{nan10a}, which are an updated version of the BH91 relations, with the new mean metallicity for M81 GCs.  The position of M81 on the graph is consistent with the general trend for other galaxies.  

The mean metallicity of the M81 GCs is higher than the means of the Milky Way and M31.  This is likely due to our sample being biased toward inner, MR bulge and disk clusters due to the limited extent of the HST I-band survey.  A plot of the locations of MP and MR GCs is shown in Figure 9.  Since our objects were chosen from HST I-band images that did not extend out to the halo, we have only a few objects projected at large distances from M81, and nearly all were observed by PBH, SBKHP, or BH91.  However, the objects outside the M81 disk do appear to be more likely to be MP: only 6 MR clusters are projected outside or at the edge of the schematic disk, as compared to 11 MP clusters, a ratio of almost 2:1 in favor of MP clusters.  These findings are consistent with our knowledge of the Milky Way \citep{har01} and M31 \citep{lee08} GCSs, in which objects at large distances from the center are preferentially MP.  Figure 10 shows the metallicity of M81 GCs as a function of projected galactocentric distance for our objects and for ``Literature'' objects, which are confirmed GCs from PBH, SBKHP, and BH91.  Also shown in Figure 10 is a metallicity gradient, a $\sigma^{-2}$-weighted linear least-squares fit to metallicity as a function of galactocentric radius.  The mean metallicity appears to remain nearly constant within $\sim$ 7 arcmin, and becomes increasingly dominated by MP clusters at larger distances, creating a slight overall metallicity gradient of $-0.025 \pm 0.020$ dex kpc$^{-1}$.  Eliminating the eight starlike clusters from the sample, the metallicity gradient increases to $-0.034 \pm 0.020$ dex kpc$^{-1}$.

Figure 10 also suggests that there may be a metallicity gradient among the MP clusters alone.  A least-squares linear fit to the MP GCs alone gives a slope of $-0.031 \pm 0.012$ dex kpc$^{-1}$.  No clear evidence of a metallicity gradient is seen in the MR subsample, but large uncertainties and sample biases make this difficult to determine.  In theory, the apparent metallicity gradient in the MP subsample could be an artifact of differences between the metallicity scales of our work and PBH, SBKHP, and BH91.  All confirmed GCs beyond $\sim$11 kpc and most confirmed GCs beyond 7 kpc are from other spectroscopic studies, primarily PBH.  However, we find that PBH/SBKHP/BH91 MP clusters alone have a metallicity gradient of $-0.045 \pm 0.014$ dex kpc$^{-1}$.  Furthermore, if we compare the mean of the metallicities of the 12 GCs we observed that were also observed by SBKHP, PBH, or BH91, we find that our mean metallicity is only 0.11 dex higher than theirs and within the scatter in the mean of the 12 metallicities ($<[Fe/H]>_{us} = -0.74 \pm 0.14$ vs. $<[Fe/H]>_{literature} = -0.83 \pm 0.20$).  The 0.11 dex mean metallicity offset is smaller than the 0.31 dex drop in metallicity over 10 kpc (half the maximum projected distance of all confirmed GCs) given by the total MP GC metallicity gradient.  It is therefore plausible that there is a genuine MP GC metallicity gradient in the M81 GCS.  Photometric evidence of similar MP metallicity gradients has been found in other large galaxies, mostly early types \citep{gei96,for01,har09a,har09b}.  A metallicity gradient in the MP subpopulation supports the notion that the MP GCs predominantly formed via in situ \citep{for97} or biased accretion \citep{bea02,str05,rho05} methods, rather than having been accreted from smaller galaxies \citep{cot98}.

Figure 11 shows the total number of GCs (MR and MP) as a function of galactocentric distance in kpc for confirmed GCs in M81, the Milky Way, and Andromeda within 20 kpc of their respective galactic centers.  The 150 Milky Way GC distances are from \citet{har96} and are true distances from the galactic center.  The 294 M31 GC distances are for objects labeled as confirmed GCs (classification number = 1, confirmation flag = 1 or 9) in \citet{gal07} and are projected distances.  The distribution of confirmed M81 clusters resembles the (absolute) distribution of Milky Way GCs in the inner regions, but with relatively few objects at large galactocentric distances.  A two-sided Kolmogorov-Smirnov test indicates a fair but inconclusive probability (P = 0.37) that the Milky Way and M81 distributions are the same within 7 kpc of their respective galactic centers.  In the Milky Way, 69\% of GCs lie within 10 kpc, 19\% lie at distances of 10-20 kpc, and 13\% lie at distances greater than 20 kpc, beyond the reach of Figure 11.  In M31, 61\% of confirmed GCs are projected within 10 kpc, 26\% are projected between 10 and 20 kpc, and 13\% are projected more than 10 kpc from the center.  In M81, however, 88\% of confirmed GCs are projected within 10 kpc of the center, and the remaining 12\% are projected between 10 and 20 kpc, with none thus far projected more than 20 kpc from the center due to lack of adequate radial coverage.  Thus, our M81 confirmed GC sample appears deficient in halo GCs compared to the more complete Milky Way and M31 samples.  As Figures 9 and 10 indicate, our M81 sample is probably deficient in MP GCs as a result of this lack of halo GCs.

To better understand how a bias toward inner clusters may be affecting our mean sample metallicity, we can compare mean metallicities of M81 GCs at different galactocentric distances to those of Milky Way GCs from \citet{har96} and M31 GCs from \citet{per02}.  Excluding about half the $R > 10$ kpc GC sample from the Milky Way and M31 reduces their mean metallicities ($-1.30$ for the Milky Way and $-1.24$ for M31) by only 0.05-0.06 dex.  If we do have half of the halo population of M81 in our sample, then perhaps the GCS of M81 is more MR than either of these two galaxies.  However, the mean metallicity of GCs in the inner 5 kpc of M81 is very similar to the mean metallicities of M31 and Milky Way GCs within this distance: $<[Fe/H]>_{R < 5 kpc}$ = $-1.03$ for M81 ($-1.05$ excluding the 8 starlike objects), $-1.00$ for M31, and $-1.07$ for the Milky Way.

\section{GC Kinematics}
\subsection{Rotation, Mean Velocity, and Velocity Dispersion}
Figure 12 shows the GC radial velocities as a function of position angle measured in degrees east of north, and Figure 13 shows the mean velocities in 30$\degr$ position angle bins.  Overplotted are the peak and outer edge H{\small{I}} rotation curves from \citet{rot75}, and a $\sigma^{-2}$-weighted least-squares fit to the rotation curve of the M81 GCs.  The formula for the rotation curve is $V_{rot,proj.} = V_{c} + V_{pr} sin (\phi - \phi_{0})$, where $V_c$ is the mean velocity, $V_{pr}$ is the projected radial velocity ($V_{rot} sin(i)$), and $\phi_{0}$ is the position angle of the rotation axis.  To deproject our rotational velocities, we adopted an inclination angle of 59$\degr$ \citep{rot75} and a position angle of 157$\degr$ \citep{rc3}.  Figure 14 shows the GC velocity vs. position angle at different distances from M81, following the distance binning of SBKHP: R $<$ 4 kpc, 4 $\leq$ R $\leq$ 8 kpc, and R $>$ 8 kpc.  Table 8 summarizes the kinematics of the GCs at different distances and metallicities.  The first column of Table 8 names the GC subpopulation.  The MR subpopulation is defined as all GCs with [Fe/H] $\geq -1.06$, and the MP subpopulation is all GCs with [Fe/H] $< -1.06$. The second column lists the number of clusters in the subpopulation.  The third column lists $V_{c}$.  The fourth column lists the deprojected rotational velocity $V_{r}$, and the fifth column lists the projected radial velocity $V_{pr} = V_{r} sin(i)$ determined by our least squares fit.  The sixth column lists the position angle $\phi_{0}$.  The final two columns list the velocity dispersion of each subpopulation with and without correction for the rotation of the subpopulation.

The mean radial velocity of all 108 GCs is -23 $\pm$ 4 km s$^{-1}$; excluding the 8 ``starlike'' objects reduces the mean radial velocity to -24 $\pm$ 4 km s$^{-1}$.  In order to compare our mean GC velocity to a velocity for the M81 nucleus derived from a relatively internally consistent set of measurements, we searched the Center for Astrophysics spectroscopy archives for spectra of the M81 nucleus from the Z-Machine and FAST spectrographs.  We found 42 spectra, 15 from the Z-Machine dating between 1978 and 1980 and 27 from FAST dating between 1994 and 1996.  All spectra had radial velocities measured via either cross-correlation (like the radial velocities of our GC spectra) or a combination of cross-correlation and emission lines.  Most also had heliocentric velocity corrections listed.  For those that did not have heliocentric corrections - the older FAST spectra - we calculated corrections using the IRAF task ``bcvcorr'' in the rvsao package.  We applied the heliocentric corrections to each velocity and found an error-weighted mean of -21 $\pm$ 3 km s$^{-1}$ --- highly consistent with our mean GC radial velocity.

The GCS of M81 shows strong evidence of rotation as a whole, at a rate of 108 $\pm$ 22 km s$^{-1}$, roughly half the \citet{rot75} peak H{\small{I}} rotation rate of $\sim$240 km s$^{-1}$. This rotation rate is only slightly less than that of the {\it{MR}} subpopulation of the Milky Way (118 km s$^{-1}$, Harris 2001), and larger than that of the Milky Way as a whole, yet is modest compared to the rapid rotation rate of 190 km s$^{-1}$ that \citet{lee08} find for the M31 GCS.   The exclusion of the 8 starlike objects raises the total M81 GC rotation rate by a statistically insignifcant amount, to 113 $\pm$ 23 km s$^{-1}$.  The rotation-uncorrected velocity dispersion of the M81 GCS is 145 km s$^{-1}$, and is unaffected if the 8 starlike objects are excluded.  Applying the rotation correction gives a velocity dispersion of 130 km s$^{-1}$, similar to that of M31 \citep{lee08}.  Our high overall rotation rate may be attributable to the bias toward inner, bulge and disk GCs in our sample discussed in \S 4.  We discuss the rotation rates of subpopulations defined according to projected distance and metallicity below.

In Figure 14 and Table 8, rotation is visibly apparent in the low and intermediate projected distance bins, but less obvious for GCs at large projected distances.  The inner sample yields a (deprojected) rotation velocity of 133 $\pm$ 40 km s$^{-1}$, while the middle sample yields a rotation velocity of 87 $\pm$ 24 km s$^{-1}$.  This contrasts somewhat with the results of SBKHP, who find the greatest rotation to occur among the intermediate-projected-distance GCs.  The outer sample nominally has a higher rotational velocity, 101 $\pm$ 33 km s$^{-1}$, but it is more uncertain.  The velocity dispersion of the outer clusters is reduced by only 1 km s$^{-1}$ when this rotation is applied, as opposed to 4 km s$^{-1}$ for the intermediate-projected-distance population and 17 km s$^{-1}$ for the inner population.  At intermediate projected distances, the rotation-corrected velocity dispersion is 133 km s$^{-1}$, and at large projected distances it is 132 km s$^{-1}$, similar to the 130 km s$^{-1}$ calculated for the entire GCS.  At small projected distances, the rotation-corrected velocity dispersion is slightly higher, at 138 km s$^{-1}$.

The difference between our results and SBKHP's results may be due to SBKHP having a much smaller sample at R $<$ 4 kpc (18 clusters as opposed to 54), thereby not making the internal rotation pattern as clear as with our much larger inner GC sample.  We also checked the possibility that background light from M81, which on average would resemble an old stellar population such as a GC, is contaminating our GC velocities and metallicities.  We found 6 sky-offset spectra (5 of them corresponding to the locations of GCs and one to the location of an H{\small{II}} region, all less than 1.6$\arcmin$ from the center of M81) in which the M81 background light from the bulge was strong enough to extract and analyze separately as if it were a star cluster.  The metallicities derived for these background bulge light samples were all around -0.4 dex with little scatter (0.04 dex), whereas the mean metallicity of our GCs is -0.96 dex with a scatter of 0.65 dex.  Even the very innermost GC sample, within 2 kpc of the center of M81 and thus most likely to be polluted by M81 background light, has a mean metallicity of -0.94 dex and a scatter of 0.59 dex.  We therefore conclude that our inner GC spectra are not simply M81 background light pollution, and thus our velocities and metallicities are most likely valid.

Figure 15 shows the velocity as a function of position angle and the rotation curve fits for the MR and MP subpopulations.  The MP subpopulation has one significant outlying cluster, 1352 (marked with an ``x'' in Figure 12).  It is a very bright (I = 16.83) cluster located about 0.66$\arcmin$ from the nucleus, with a radial velocity of 353 $\pm$ 12 km s$^{-1}$.  If this object is included in the least-squares weighted fit to the MP rotation curve, the MP GC rotational velocity is increased to a large value (161 km s$^{-1}$), and correcting for the rotation actually increases the velocity dispersion from 142 km s$^{-1}$ to 154 km s$^{-1}$.  We find the best rotational velocity fit, giving the greatest reduction in MP GC velocity dispersion, by using ${\sigma}^{-2}$ weights and excluding Object 1352.  MR GCs rotate at about 122 km s$^{-1}$, while MP GCs rotate at about 67 km s$^{-1}$ if they rotate at all.  The MP velocity dispersion is reduced by only 1 km s$^{-1}$ when the rotation correction is applied, and the uncertainty in the rotation velocity is quite large (38 km s$^{-1}$, nearly half the magnitude of the rotation velocity itself).  This is similar to what is found in the Milky Way \citep{zin85, har01}, where inner, MR bulge/disk GCs rotate significantly but the MP halo population as a whole does not.

We can also revisit the effects of Object 1352 on the total and inner M81 GC rotation rates.  Recalculating the total rotation curve without Object 1352 gives V$_{r,tot}$ = 93 $\pm$ 17 km s$^{-1}$ (down from 108 km s$^{-1}$), V$_0$ = -29 $\pm$ 11 km s$^{-1}$, and $\phi_0 = 280\degr$.  Recalculating the inner rotation curve minus Object 1352 gives V$_{r,tot}$ = 102 $\pm$ 29 km s$^{-1}$ (down from 133 km s$^{-1}$), V$_0$ = -34 $\pm$ 19 km s$^{-1}$, and $\phi_0 = 271\degr$.  We then compared the rotation-corrected velocity dispersions including and excluding Object 1352, calculated by subtracting each GC's predicted radial velocity using the above rotation curve results from its actual radial velocity and finding the dispersion of these adjusted velocities.  The total rotation-corrected velocity dispersion (including 1352) increases slightly to 131 km s$^{-1}$, and the inner rotation-corrected velocity dispersion (including 1352) is slightly improved at 138 km s$^{-1}$.  Thus, Object 1352 increases the rotation velocities in these subsamples by about 1 $\sigma$, but does not have a drastic effect on the velocity dispersion as with the MP GCs.

Overall, the analysis of rotation rate as a function of distance and metallicity suggests that, while there may be some rotation among outer GCs and possibly MP GCs, the most obvious rotation is among inner and MR GC populations, much as in the Milky Way.  Since GCs within the inner 4 kpc make up half our sample, and MR GCs within the inner 4 kpc make up a full quarter of our sample, the overall rotation rate we observe is most likely biased toward high values due to the overrepresentation of M81's rapidly rotating bulge/disk GC population.  However, unlike in the Milky Way, there is some evidence for relatively weak rotation among GCs at distances greater than 4 kpc and possibly (though this is less clear) among MP GCs.  More MP halo GCs in M81 would need to be spectroscopically confirmed, from a survey covering larger galactocentric distances, to better understand the kinematics of M81's GCS.

\subsection{Mass Estimates}
The velocity dispersion of all 108 GCs, corrected for the mean uncertainty in velocity, is 145 km s$^{-1}$.  Using the projected mass estimator \citep{hei85}, assuming isotropic orbits and subtracting a systemic velocity of -34 km s$^{-1}$ from all GC velocities, the 108 clusters yield a mass of (1.64$\pm$0.28)$\times 10^{11}$ M$_{\odot}$ for M81.  The 100 non-starlike clusters yield a total projected mass of (1.56 $\pm$ 0.27)$\times 10^{11}$ M$_{\odot}$. Errors are estimated using the bootstrap method, randomly resampling from the original set of v$^2$r values 10,000 times.  Our total M81 mass is considerably smaller than previous determinations ($\sim$4 $\times 10^{11} M_{\odot}$ in SBKHP and PBH), but this seems to be because most of our clusters come from the inner regions, leading to smaller v$^2$r values on average.  Our mass is similar to the \citet{rot75} total mass estimate of 1.7 $\times 10^{11}$ M$_{\odot}$.  Mass estimates as a function of maximum distance from the center are shown in Table 9.  The first column lists the number of clusters in each subsample defined by a maximum galactocentric distance.  The second column lists the maximum galactocentric distance of each subsample.  The third column lists the projected mass estimate derived from all clusters within that projected distance.

\section{Summary}
Overall, we find the GCS of M81 to be similar to those of M31 and the Milky Way.  A KMM test provides modest statistical evidence that the M81 GC system has a bimodal metallicity distribution, with peaks in the metallicity distribution similar to those of M31 and the Milky Way.  The mean metallicity of 107 GCs is --1.06 $\pm$ 0.07, which is higher than those of M31 or the Milky Way.  The high mean metallicity is most likely due to our sample being artificially biased toward inner, MR clusters.  There is evidence of a metallicity gradient in the GCS as a whole, and in the MP GC subsample as well.  The GCS as a whole appears to be rotating, with the strongest rotation found among MR GCs (as in SBKHP) and GCs in the inner 4 kpc, in contrast to SBKHP who found the most rotation in the 4-8 kpc range.  MP GCs seem to be rotating at a slower rate than MR GCs if at all.  This pattern of strong rotation among inner and MR clusters and relatively weak or no rotation among outer and MP clusters is reminiscent of the Milky Way.

Future studies could improve both our understanding of the MP and MR subpopulations, and of the global properties of the M81 GCS as a whole.  Background subtraction is a relative weakness of this study that affects inner GCs most strongly, and therefore long-slit individual spectroscopy on known clusters projected onto the M81 disk might help confirm or deny our results for strong rotation of the innermost GC populations and MR clusters.  A more comprehensive study of the MP halo cluster system, perhaps done with a multi-fiber spectrograph, would help to confirm the rotation or lack thereof among the outer and MP subpopulations, and better constrain the overall rotation of the M81 GCS.

\begin{figure}
\plotone{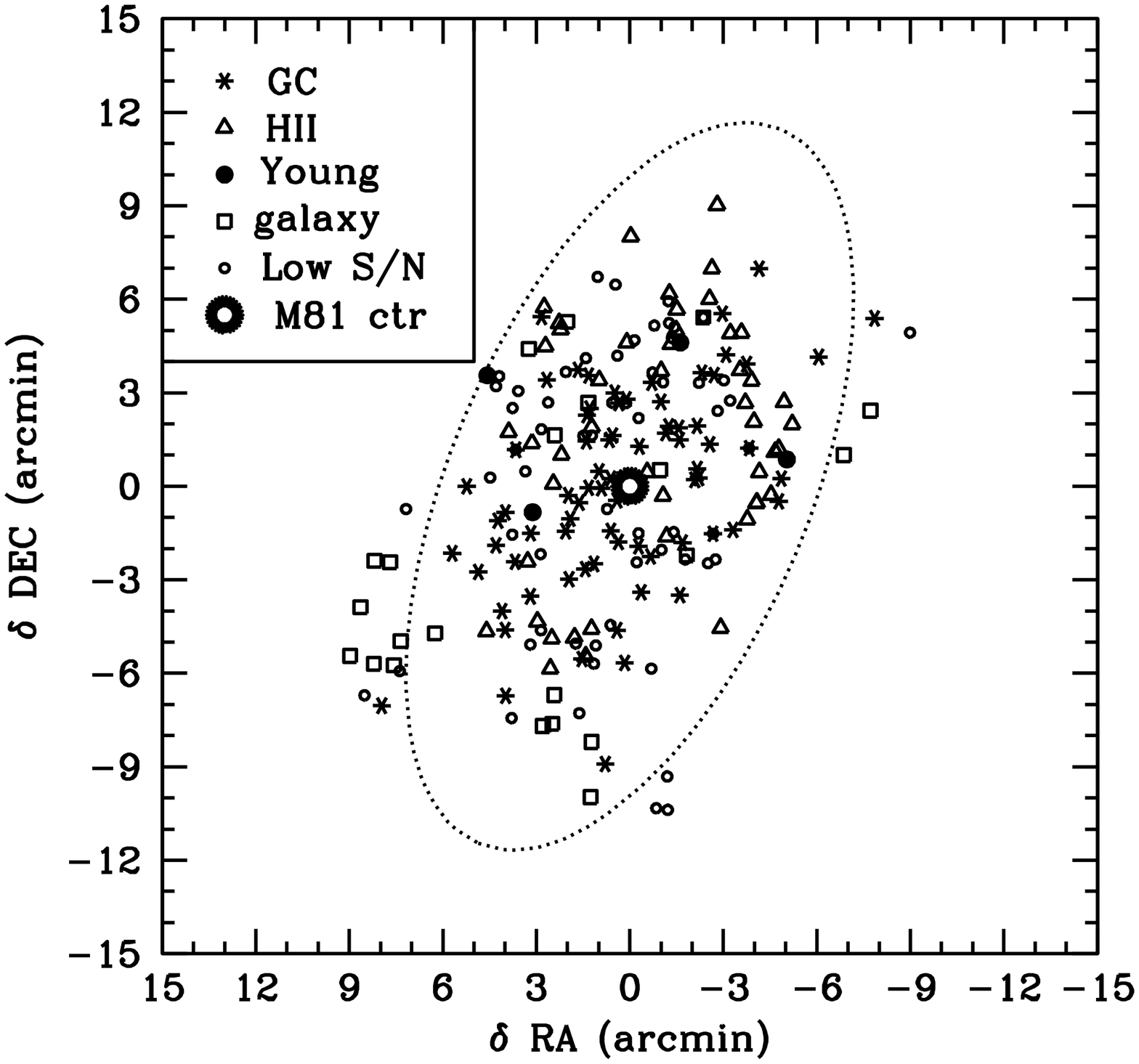}
\caption{Locations of spectroscopically observed objects.  The dotted line represents the $\mu_B$ = 25 mag arcsec$^{-2}$ isophote.}
\end{figure}

\begin{figure}
\plotone{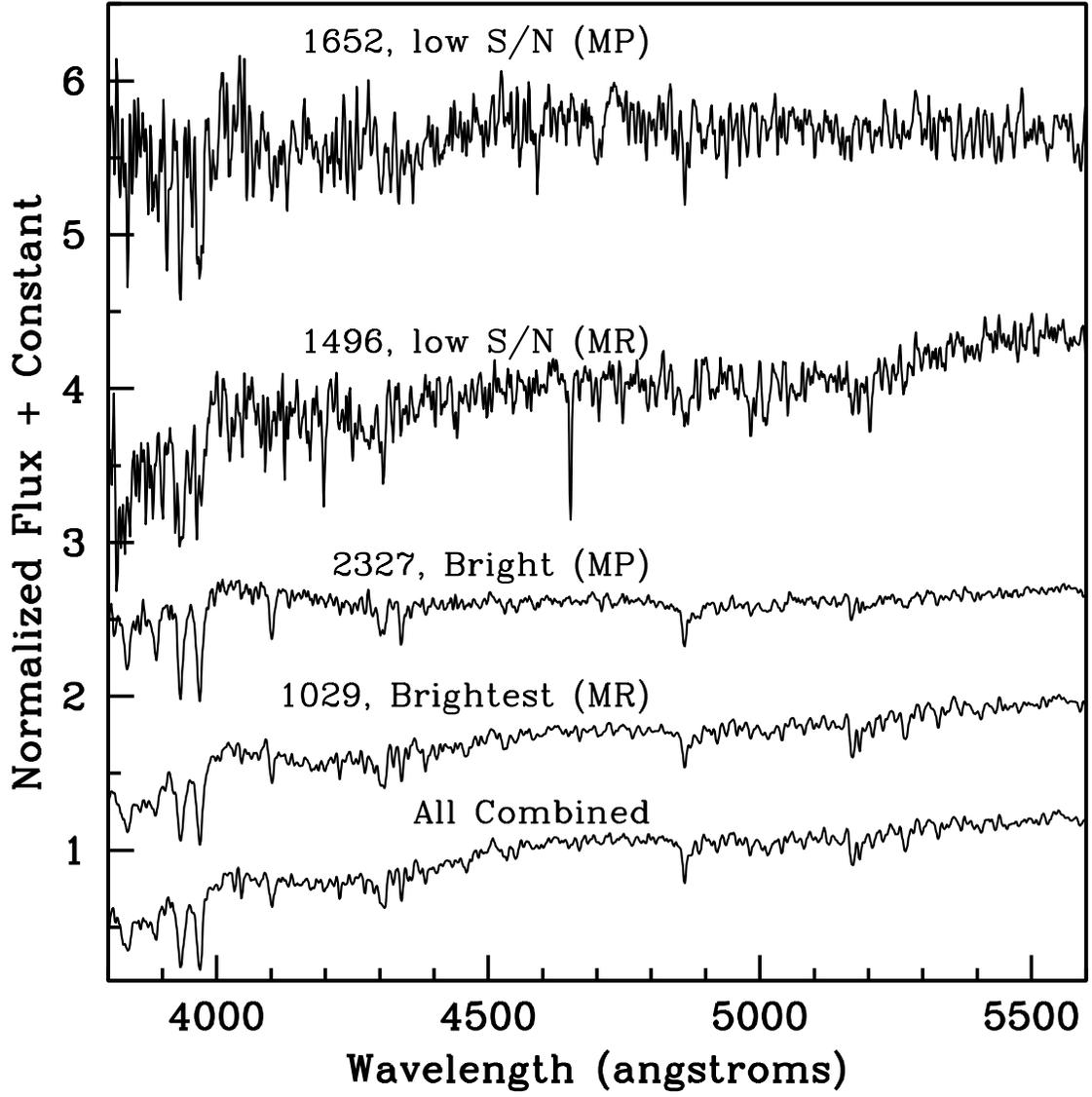}
\caption{Spectra of selected M81 GCs.}
\end{figure}

\begin{figure}
\plotone{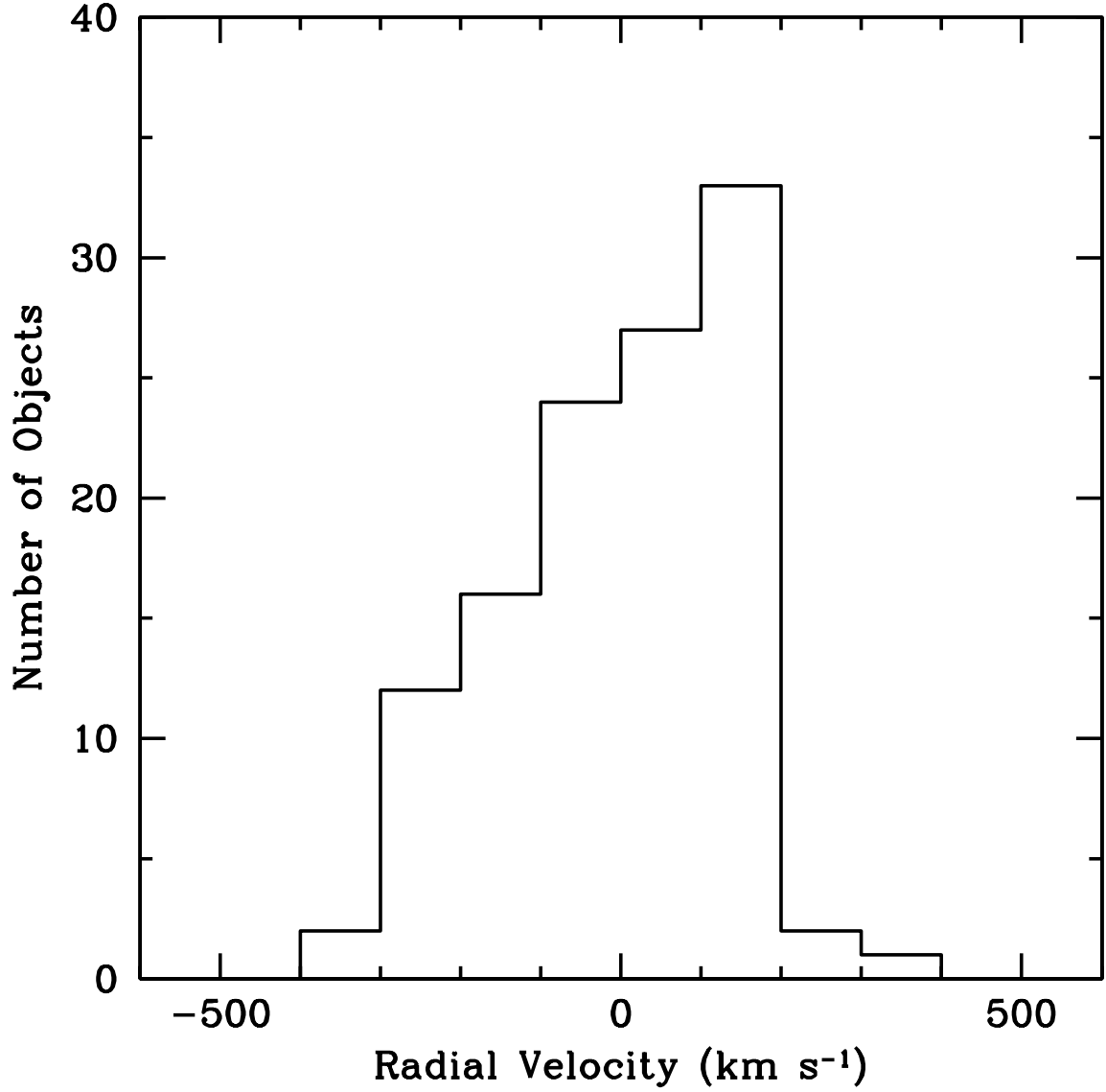}
\caption{Velocity histogram of all M81 objects (GCs, H{\small{II}} regions, and open clusters).}
\end{figure}

\begin{figure}
\plotone{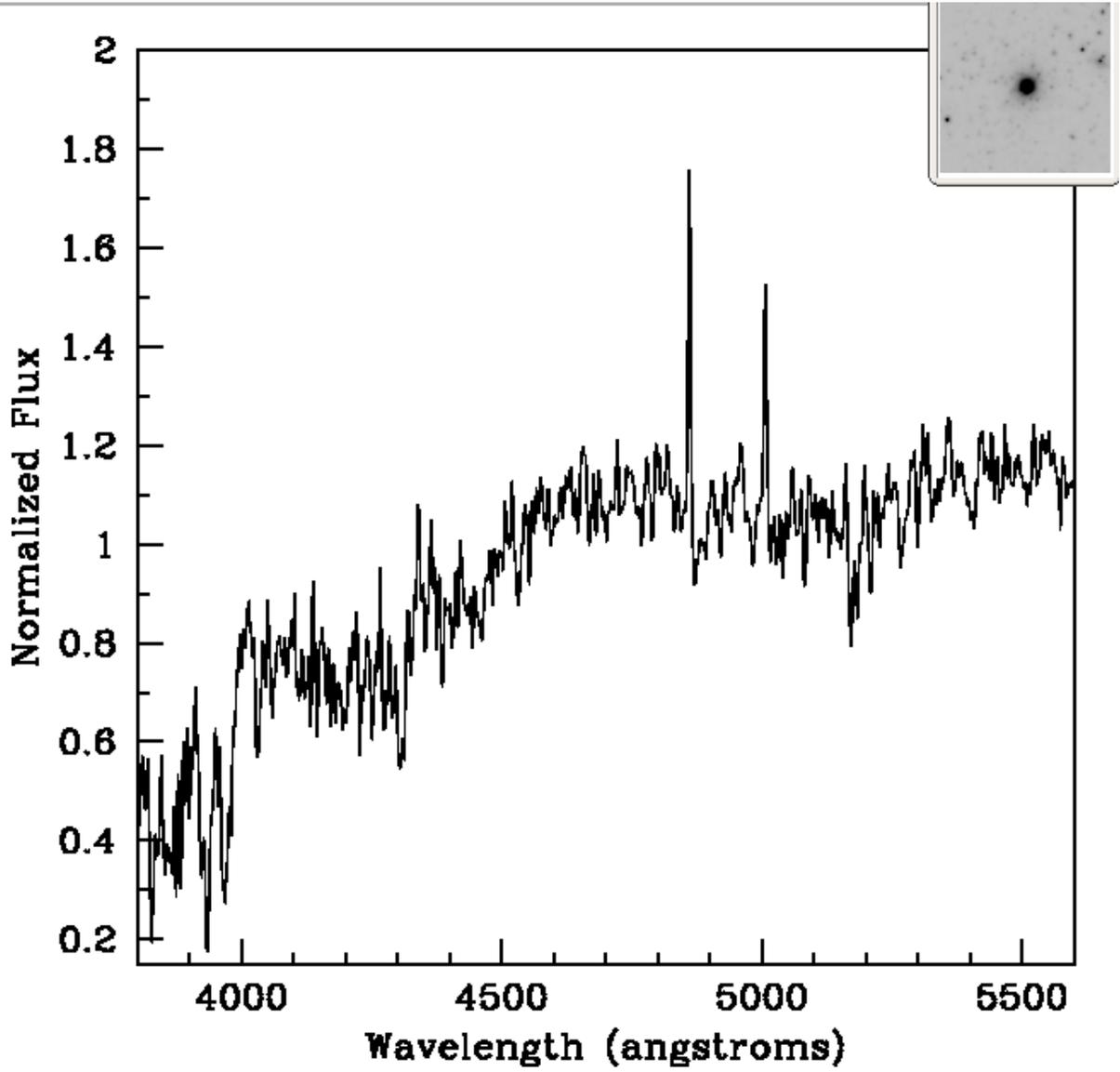}
\caption{Spectrum and HST image of Object 1859, an apparent M81 GC mixed with H{\small{II}} emission.}
\end{figure}

\begin{figure}
\plotone{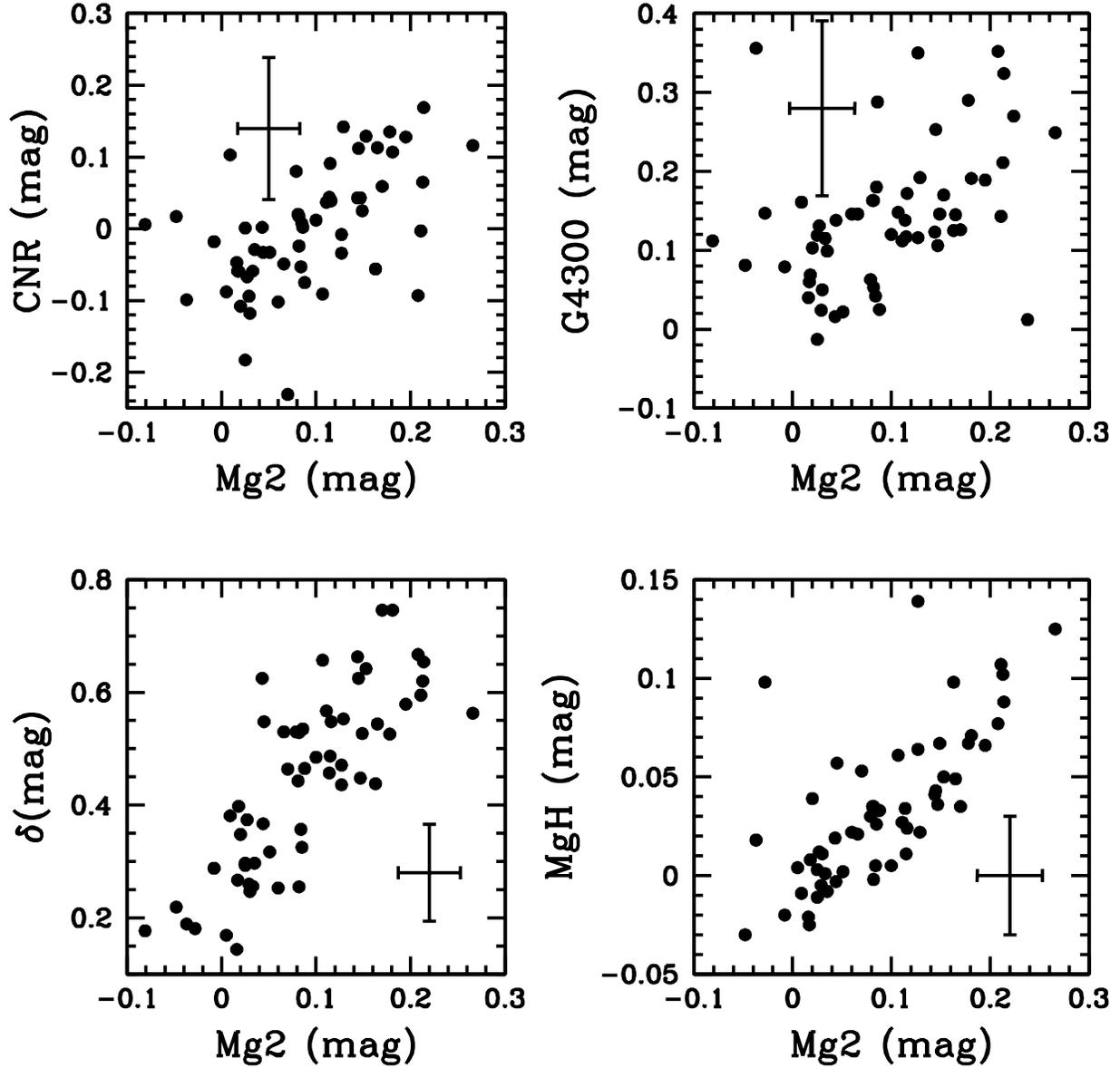}
\caption{Blue indices and MgH vs. Mg2 for M81 GCs.  Error bars shown are mean uncertainties in all M81 GC indices.}
\end{figure}

\begin{figure}
\plotone{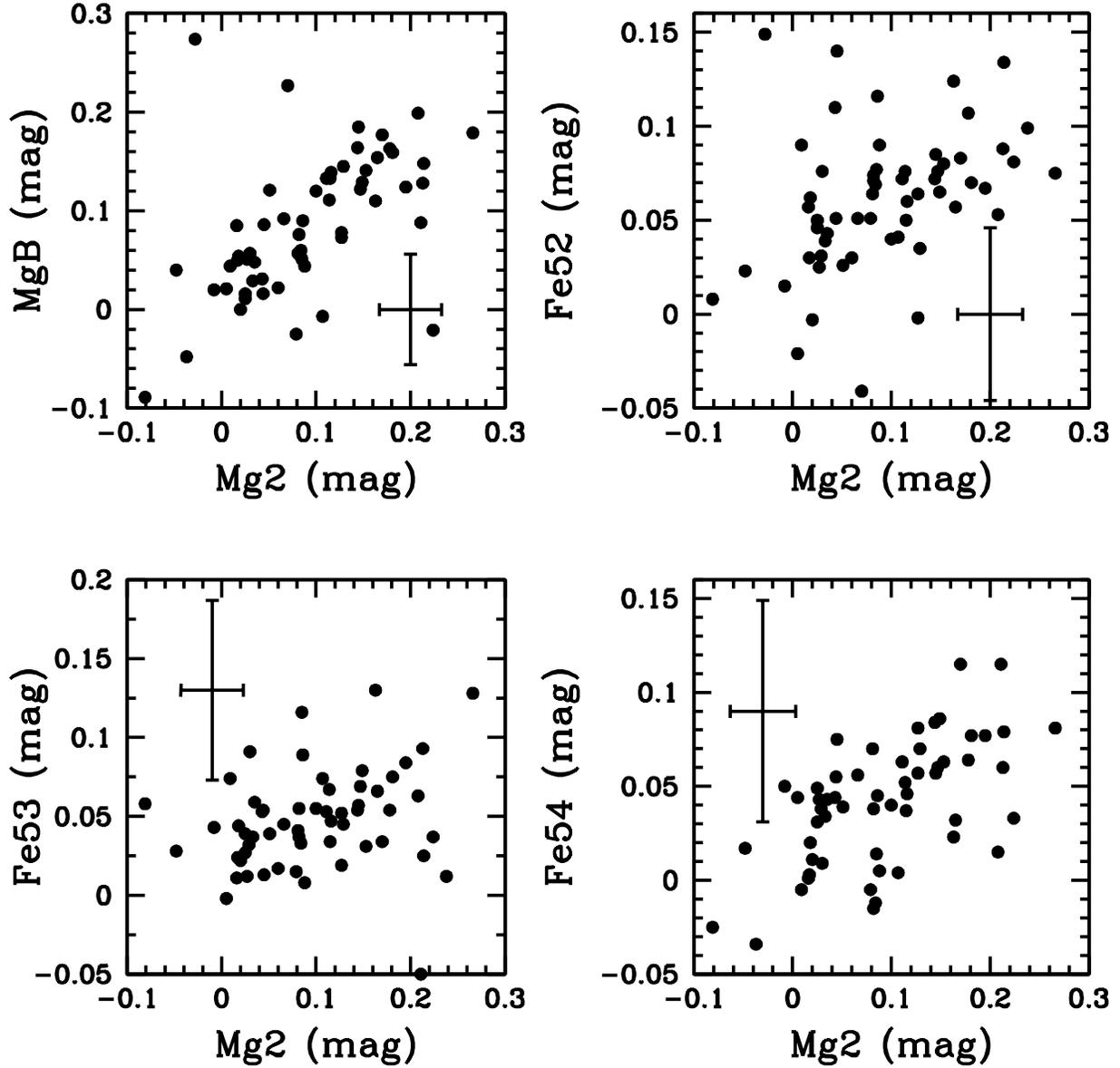}
\caption{MgB and iron lines vs. Mg2 for M81 GCs.  Error bars shown are mean uncertainties in all M81 GC indices.}
\end{figure}

\begin{figure}
\plotone{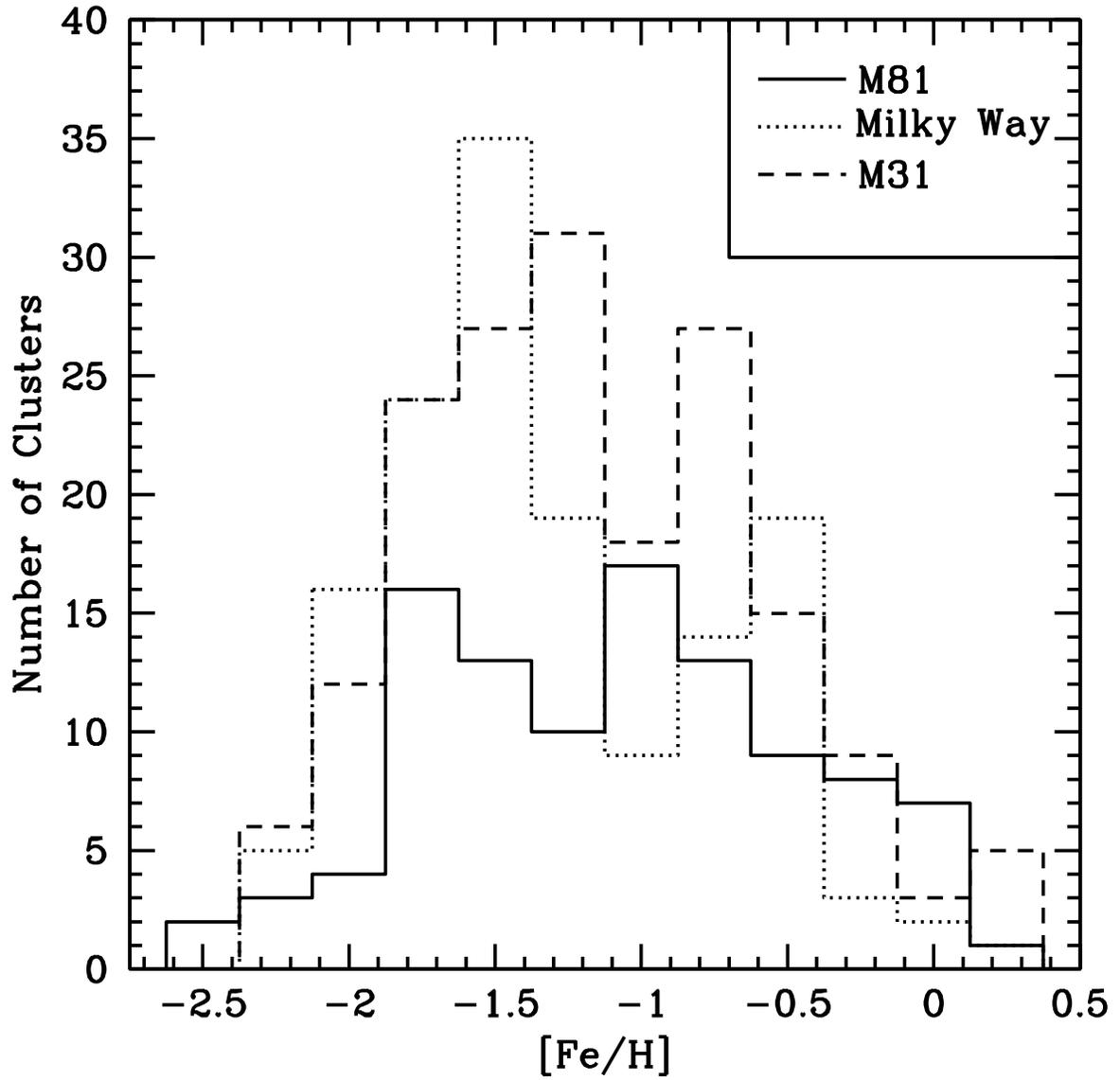}
\caption{Histogram of the metallicities of 103 M81 GCs, along with the Milky Way and M31 for comparison.}
\end{figure}

\begin{figure}
\plottwo{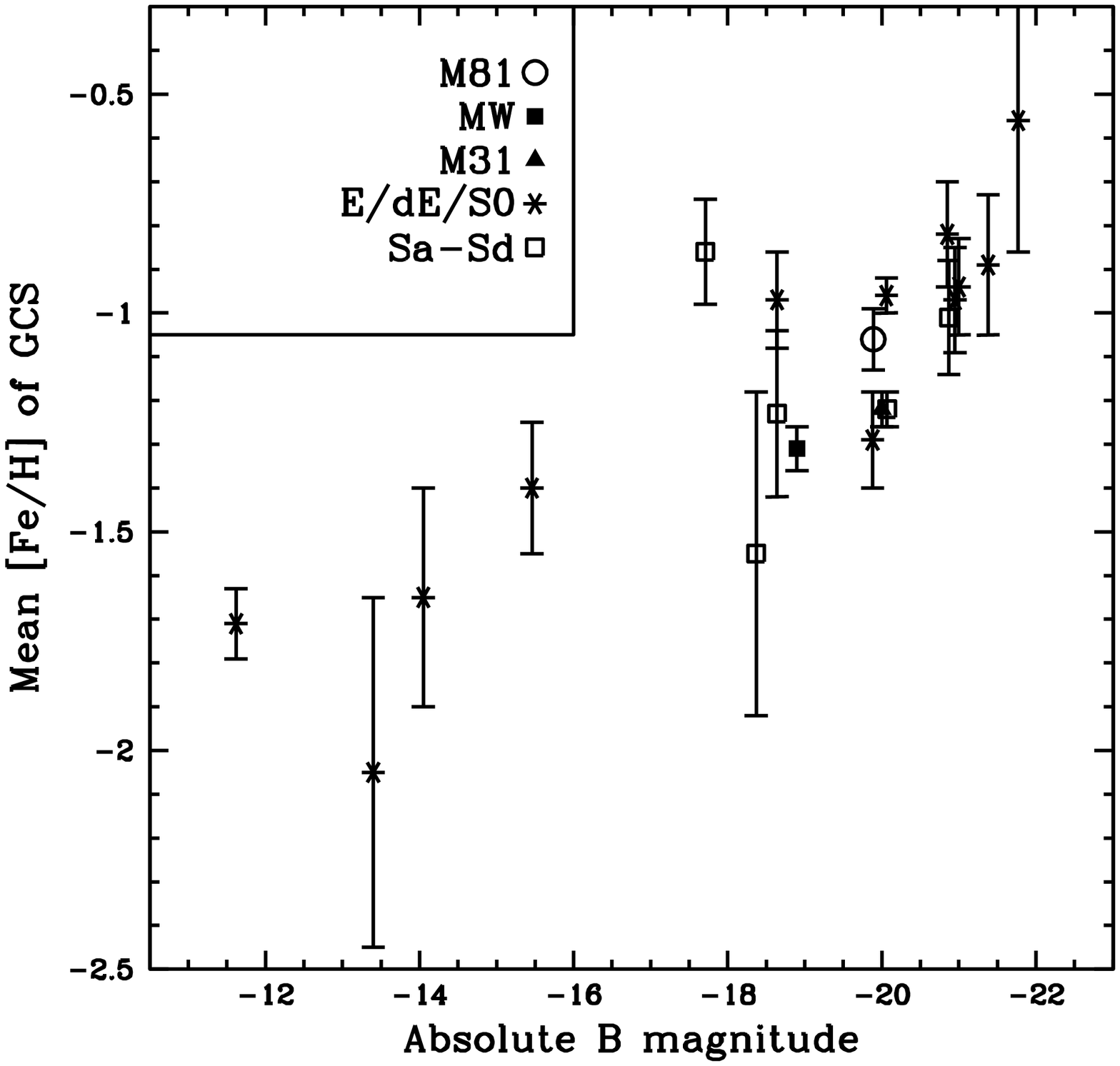}{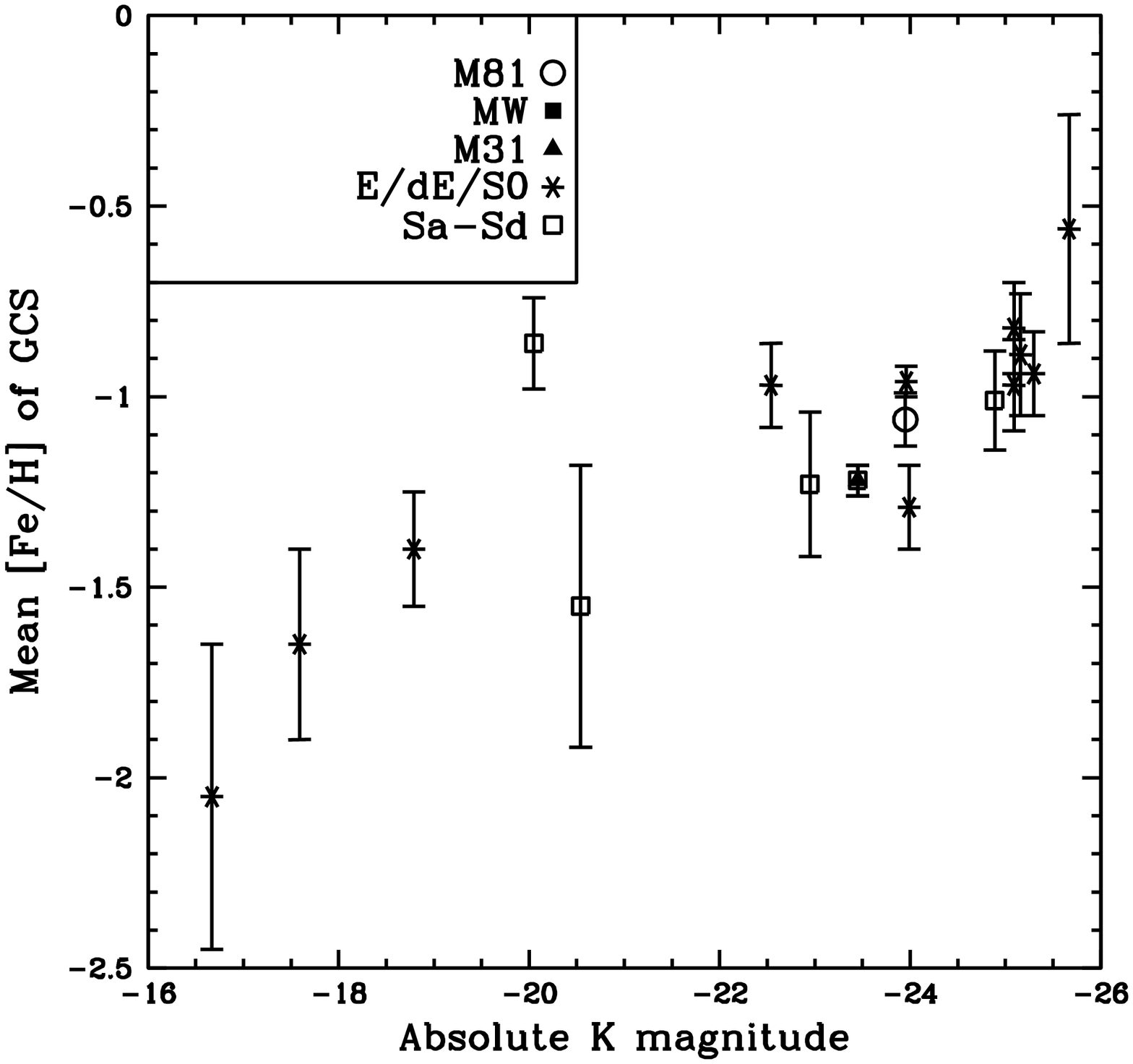}
\caption{GC metallicity-galaxy luminosity relations with our current M81 mean metallicity.}
\end{figure}

\clearpage

\begin{figure}
\plotone{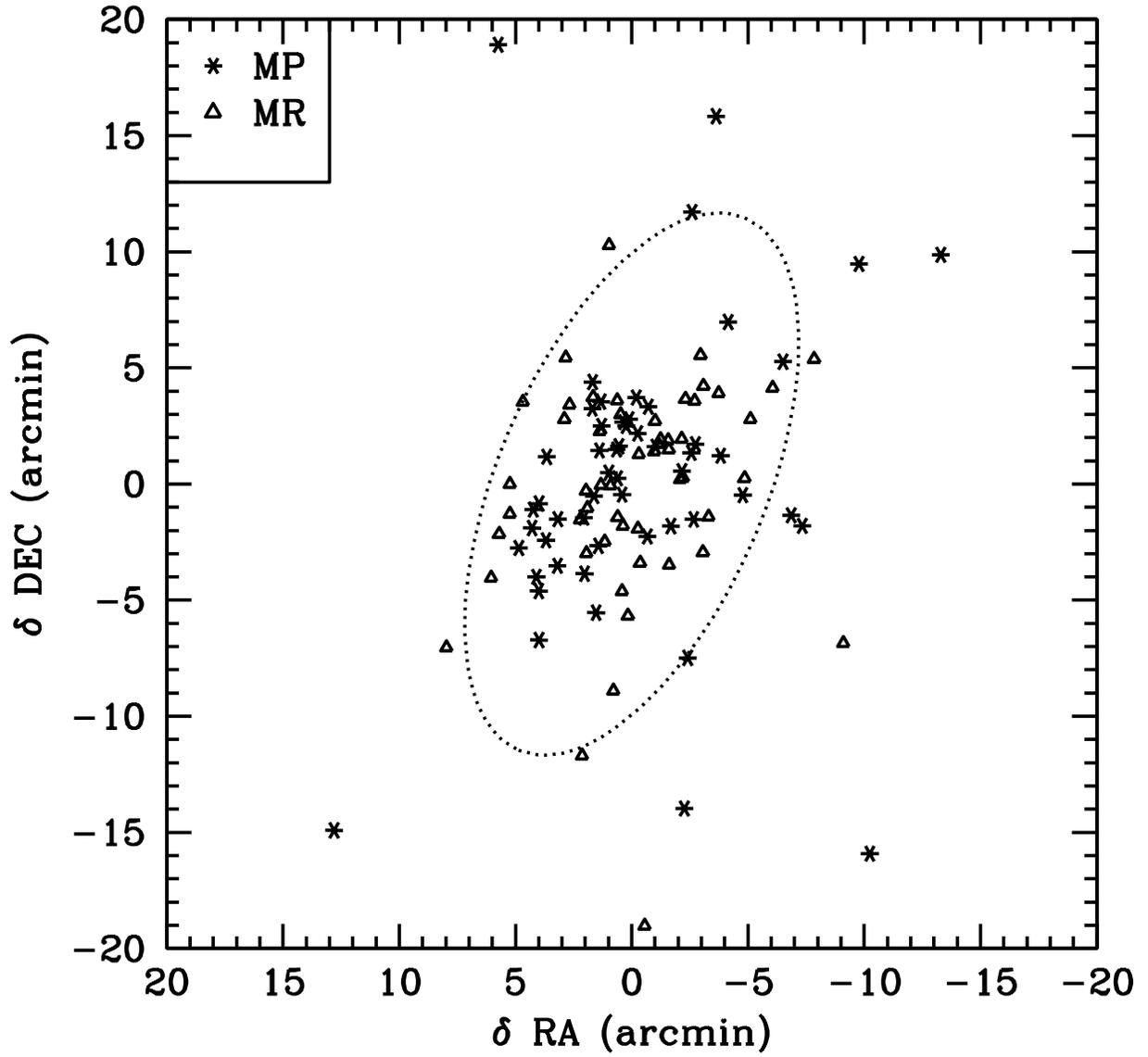}
\caption{Locations of MP and MR M81 GCs.  The dotted line represents the $\mu_B$ = 25 mag arcsec$^{-2}$ isophote.}
\end{figure}

\begin{figure}
\plotone{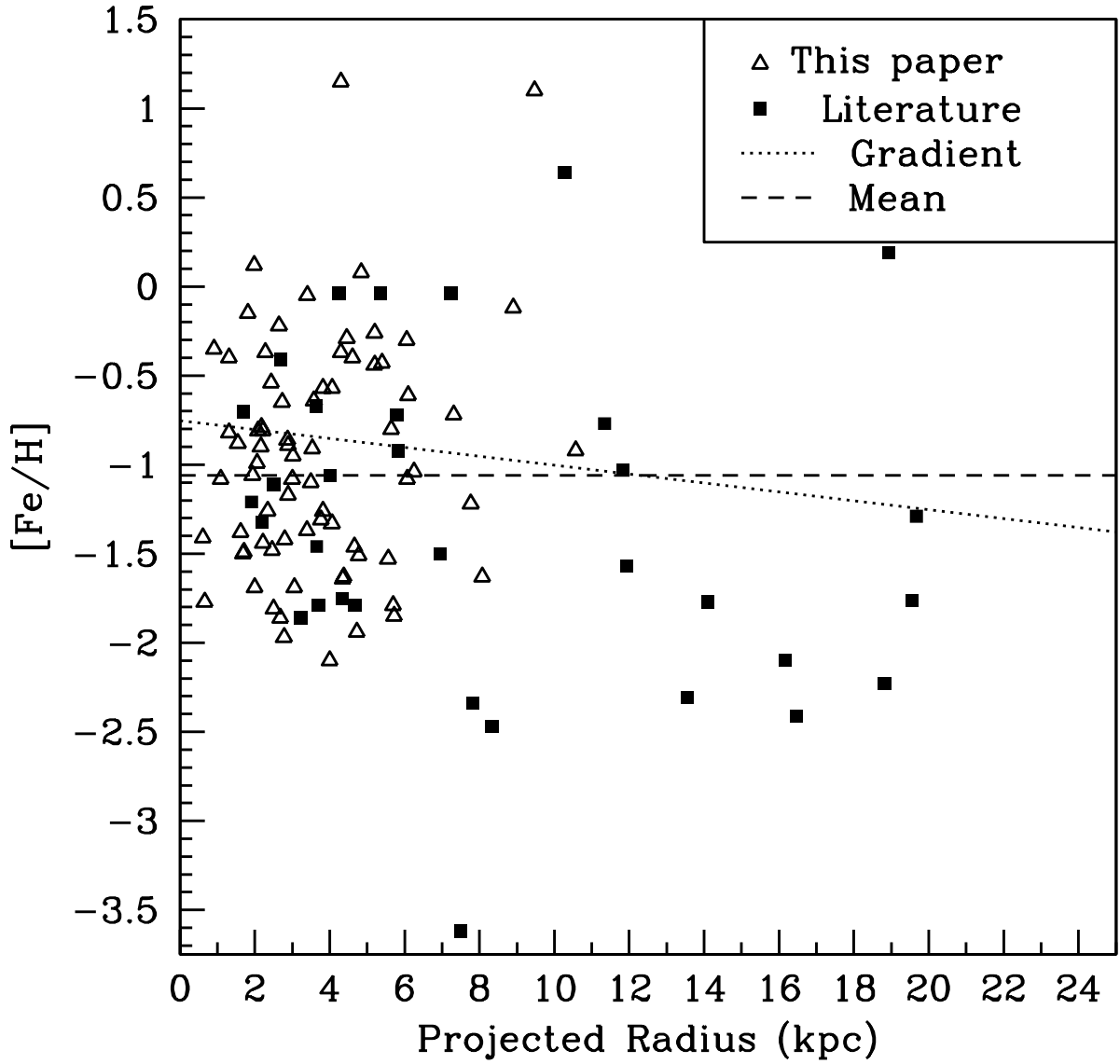}
\caption{Metallicity of GCs as a function of distance from the center of M81.  ``Literature'' objects are from PBH, BH91, and SBKHP.  The metallicity gradient is a least-squares weighted linear fit to metallicity as a function of galactocentric distance.}
\end{figure}

\begin{figure}
\plotone{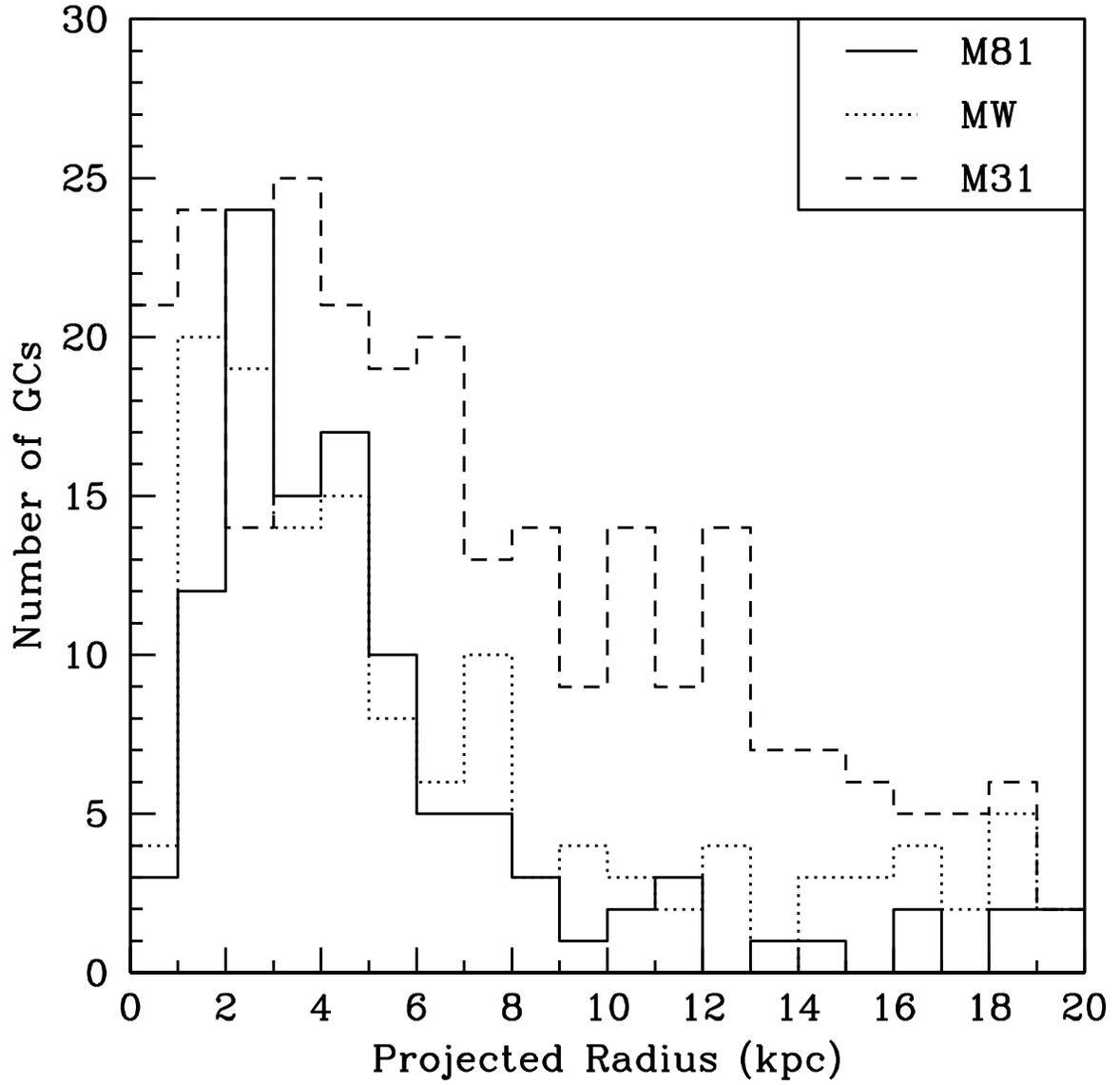}
\caption{Number of confirmed GCs in 1 kpc bins as a function of projected radius in M81 (108 confirmed GCs), the Milky Way (150 confirmed GCs), and Andromeda (294 confirmed GCs).}
\end{figure}

\begin{figure}
\plotone{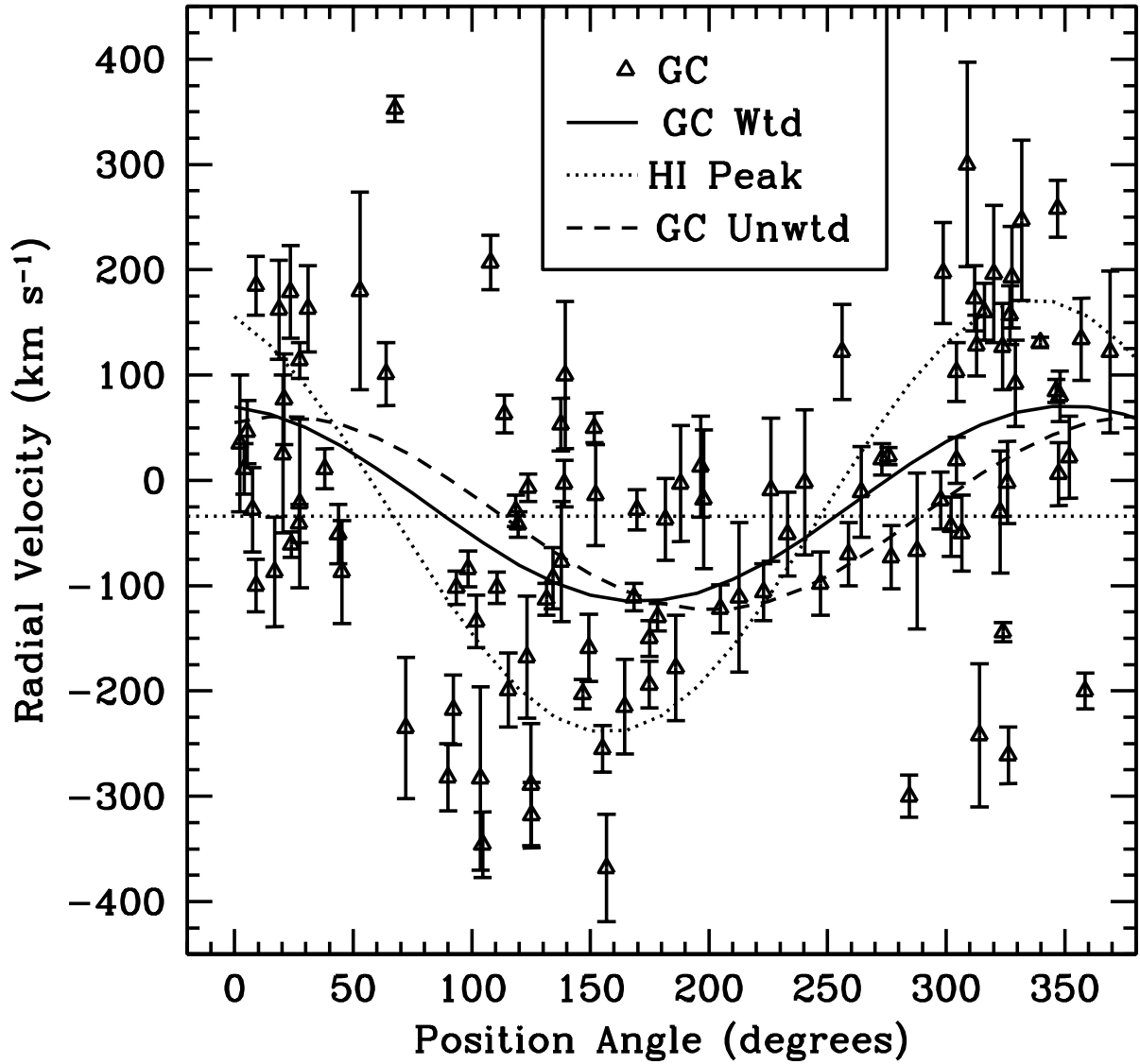}
\caption{Velocity vs. position angle for all M81 GCs, shown with the peak and outer edge H{\small{I}} rotation curves and a fit to the GC rotation.}
\end{figure}

\begin{figure}
\plotone{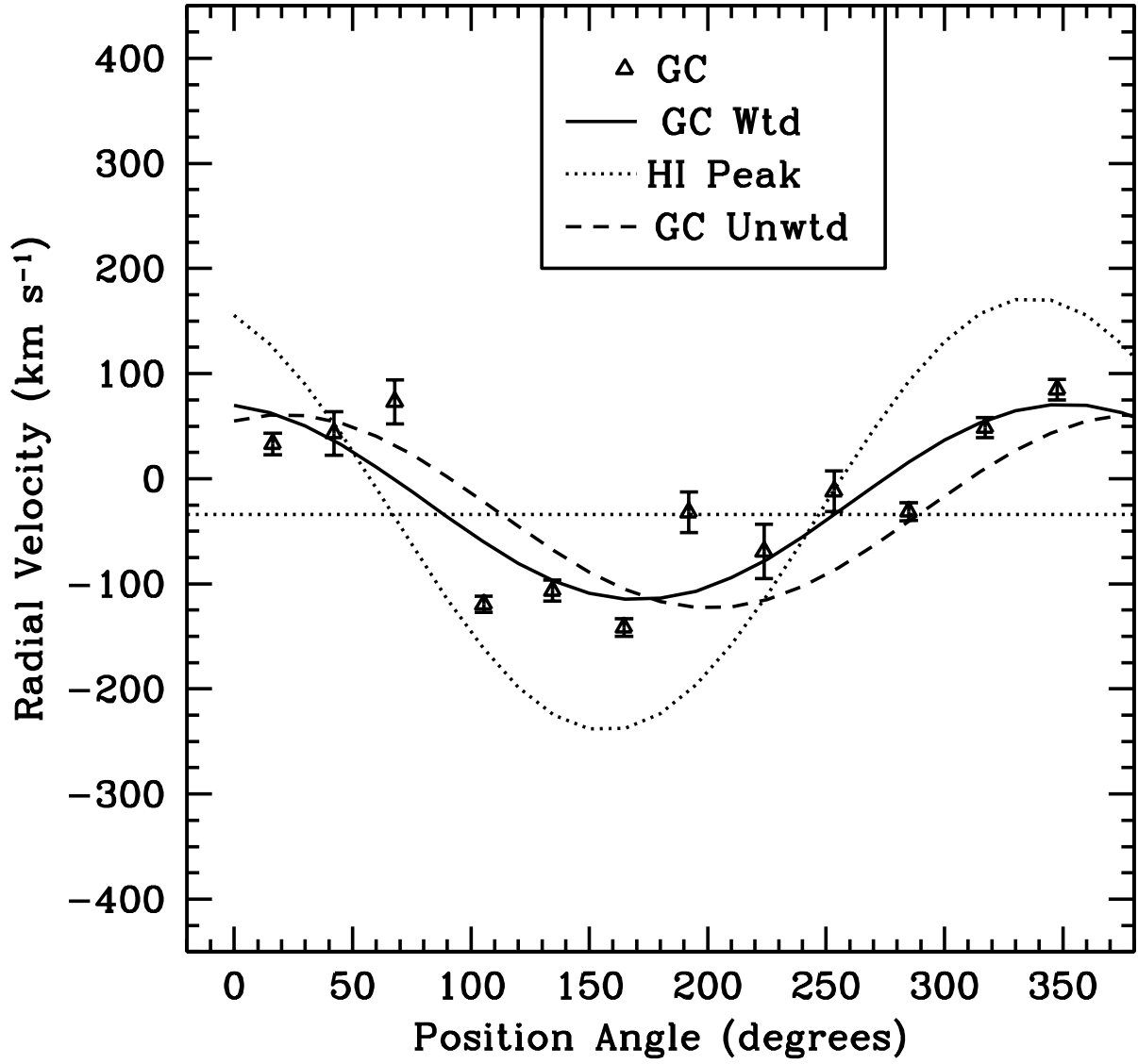}
\caption{Mean velocity vs. position angle in position angle bins of 30$\degr$.}
\end{figure}

\begin{figure}
\plotone{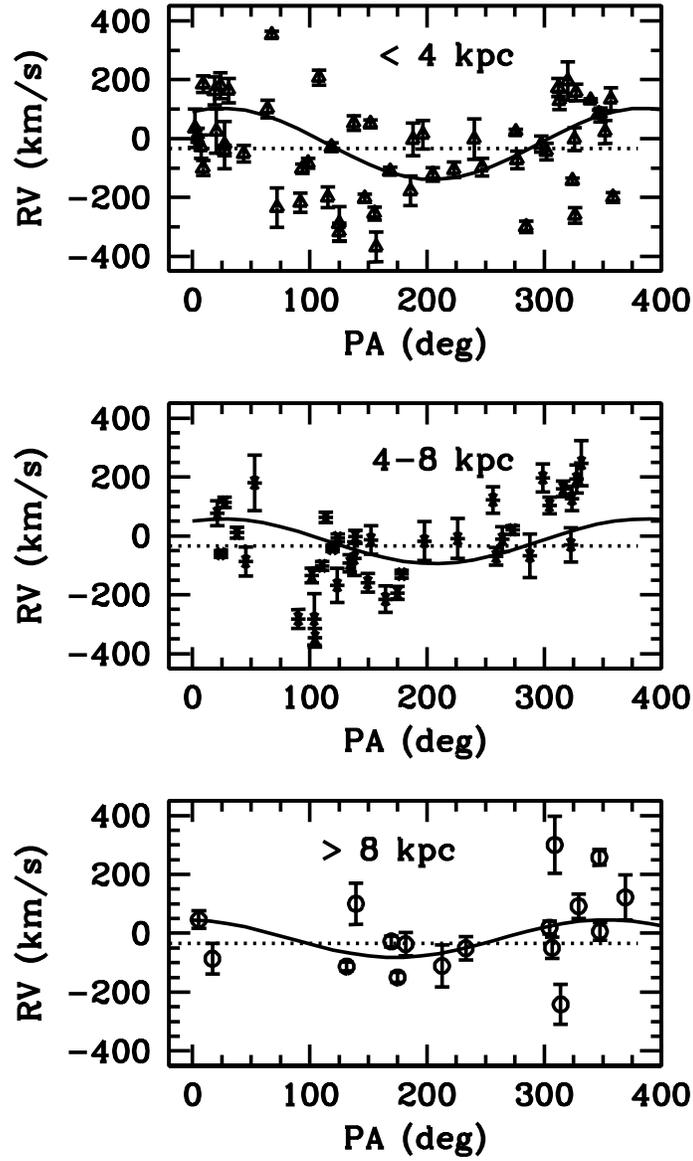}
\caption{Velocity vs. position angle for GC subsamples at various distances from the center of M81, with fits to the GC rotation.}
\end{figure}

\begin{figure}
\plotone{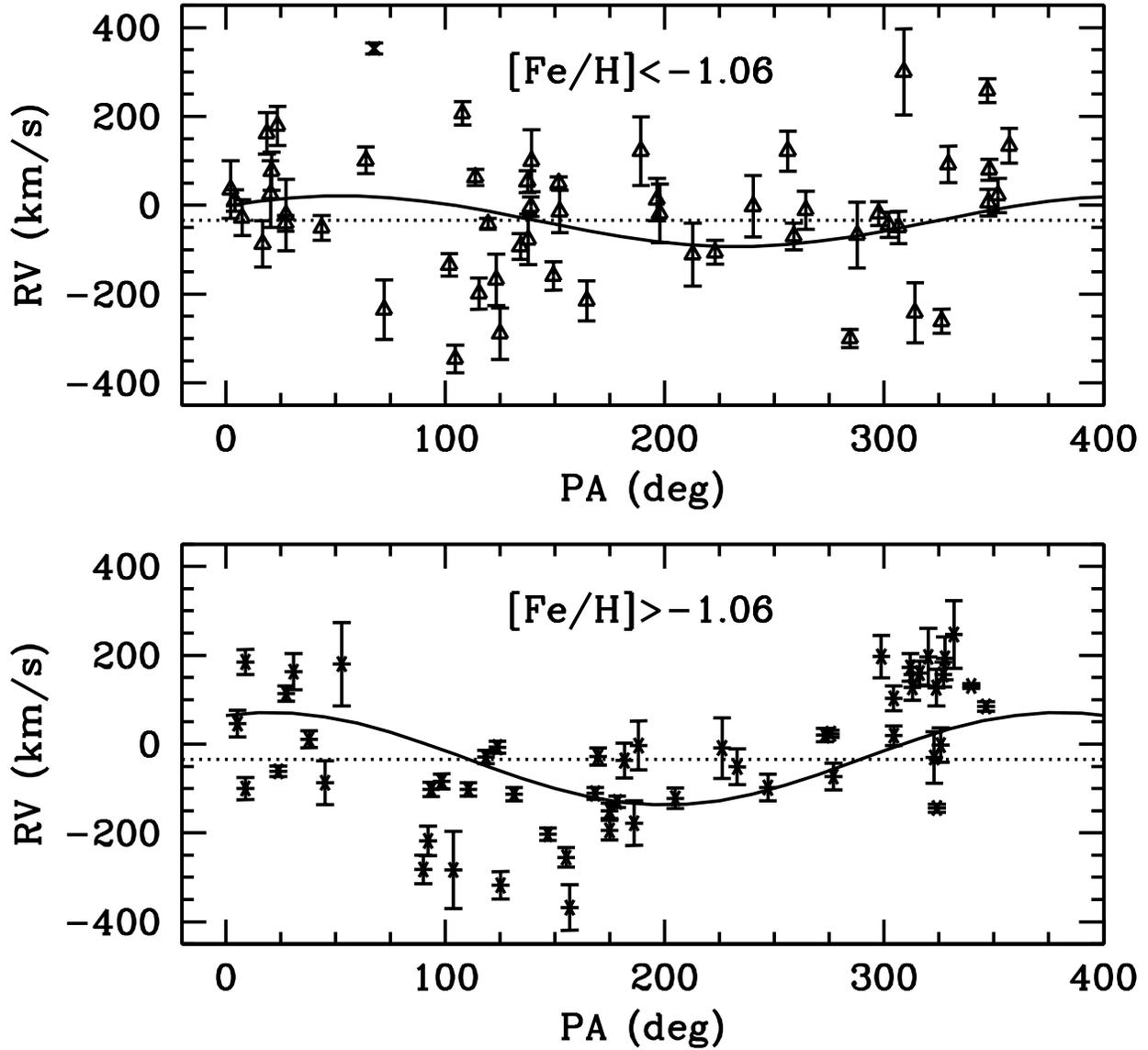}
\caption{Velocity vs. position angle for MP (top) and MR (bottom) GC subsamples, with fits to the GC rotation.  Outlying object 1352, an inner MP GC marked with an ``X'' instead of a triangle, was left out of the MP rotation fit.}
\end{figure}

\begin{deluxetable}{ccccc}




\tablecaption{Observation Information}

\tablehead{\colhead{Date} & \colhead{N$_{objects}$} & \colhead{N$_{objexp}$} & \colhead{N$_{skyexp}$} & \colhead{T$_{int}$} \\ 
\colhead{} & \colhead{} & \colhead{} & \colhead{} & \colhead{} } 

\startdata
2006 May 03 & 51 & 4 & 3 & 80 \\
2006 May 04 & 52 & 4 & 1$^a$ & 80 \\
2006 May 05 & 51 & 3 & 3 & 60 \\
2007 Nov 13 & 72$^b$ & 6 & 6 & 120 \\
2007 Nov 17 & 72$^c$ & 2$^d$ & 3$^d$ & 40 \\
\enddata


\tablecomments{(a) Three sky offsets rejected due to sky flux inconsistent with the mean sky flux of the object exposures.  Using these sky offsets results in severely over- or under-subtracted sky continuum.
(b) 21 objects are repeats of low signal-to-noise spectra from the May 2006 observations.
(c) The same objects were observed on November 13 and November 17.
(d) Four object exposures and three sky exposures rejected due to clouds.
}


\end{deluxetable}
\begin{deluxetable}{ccccccccc}


\tabletypesize{\scriptsize}


\tablecaption{GC Positions and Velocities}


\tablehead{\colhead{ID} & \colhead{Prev. spec} & \colhead{PR95 ID} & \colhead{CFT01 ID} & \colhead{RA} & \colhead{Dec} & \colhead{Dist.} & \colhead{RV} & \colhead{$\sigma_{RV}$} \\ 
\colhead{} & \colhead{} & \colhead{} & \colhead{} & \colhead{(hours)} & \colhead{(deg)} & \colhead{(arcmin)} & \colhead{(km s$^{-1}$)} & \colhead{(km s$^{-1}$)} } 

\startdata
34 & -- & 51040 & -- & 09:54:04.910 & 69:09:18.390 & 9.47 & 19 & 22 \\
108 & -- & 50934 & -- & 09:54:25.100 & 69:08:03.760 & 7.31 & 103 & 28 \\
173 & PBH 50552 & 50552 & 87 & 09:54:38.810 & 69:04:10.220 & 4.84 & 20 & 15 \\
187 & (PBH 50462) & 50462 & 97 & 09:54:39.700 & 69:03:26.460 & 4.78 & -11 & 43 \\
264 & -- & 51118 & -- & 09:54:46.470 & 69:10:53.930 & 8.08 & 92 & 41 \\
345 & -- & 50652 & -- & 09:54:50.310 & 69:05:08.410 & 4.00 & -67 & 74 \\
359 & -- & 50912 & -- & 09:54:51.050 & 69:07:50.080 & 5.40 & 160 & 27 \\
464 & (PBH 50388*) & 50388 & -- & 09:54:56.140 & 69:02:31.020 & 3.58 & -98 & 30 \\
505 & -- & -- & -- & 09:54:58.470 & 69:08:08.500 & 5.20 & 127 & 41 \\
526 & (PBH 51053) & 51053 & 30 & 09:54:59.790 & 69:09:27.370 & 6.25 & 247 & 76 \\
594 & PBH 50861 & 50861 & -- & 09:55:02.780 & 69:07:29.380 & 4.46 & -30 & 58 \\
605 & -- & 50376 & -- & 09:55:03.310 & 69:02:23.730 & 3.06 & -2 & 69 \\
628 & -- & 50659 & 41 & 09:55:04.380 & 69:05:15.820 & 2.89 & -19 & 27 \\
676 & -- & 50873 & -- & 09:55:07.260 & 69:07:34.220 & 4.30 & 193 & 48 \\
705 & -- & 50555 & -- & 09:55:08.420 & 69:04:11.130 & 2.22 & -73 & 30 \\
720 & -- & 50715 & 39 & 09:55:08.980 & 69:05:51.540 & 2.89 & 173 & 31 \\
722 & -- & 50590 & -- & 09:55:09.040 & 69:04:28.390 & 2.22 & -300 & 20 \\
743 & -- & 50548 & -- & 09:55:09.790 & 69:04:07.640 & 2.09 & 23 & 8 \\
839 & -- & 50358 & 104 & 09:55:14.280 & 69:02:06.340 & 2.46 & -106 & 27 \\
861 & -- & 50259 & -- & 09:55:15.160 & 69:00:25.800 & 3.82 & -122 & 23 \\
863 & -- & 50678 & 42 & 09:55:15.240 & 69:05:24.180 & 2.18 & 128 & 29 \\
876 & -- & 50708 & 37 & 09:55:15.610 & 69:05:47.810 & 2.44 & 196 & 65 \\
966 & -- & 50713 & 38 & 09:55:19.210 & 69:05:50.110 & 2.28 & 157 & 28 \\
993 & -- & 50697 & -- & 09:55:20.160 & 69:05:37.620 & 2.06 & -2 & 39 \\
1029 & -- & 50777 & -- & 09:55:21.920 & 69:06:37.630 & 2.87 & 131 & 5 \\
1089 & SBKHP 14 & 50834 & 52 & 09:55:25.190 & 69:07:14.650 & 3.39 & 80 & 24 \\
1104 & -- & -- & 106 & 09:55:25.700 & 69:01:39.920 & 2.34 & 13 & 48 \\
1154 & -- & 50263 & -- & 09:55:29.130 & 69:00:31.010 & 3.40 & -178 & 50 \\
1162 & -- & 50655 & -- & 09:55:29.730 & 69:05:11.680 & 1.31 & 85 & 11 \\
1172 & -- & 50351 & 108 & 09:55:30.100 & 69:01:59.540 & 1.94 & -3 & 55 \\
1257 & -- & 50786 & -- & 09:55:34.410 & 69:06:42.240 & 2.78 & 35 & 65 \\
1265 & -- & 50104 & -- & 09:55:34.920 & 68:58:14.820 & 5.64 & -130 & 13 \\
1300 & -- & 50773 & -- & 09:55:37.140 & 69:06:35.530 & 2.68 & -28 & 40 \\
1301 & SBKHP 03 & 50359 & -- & 09:55:37.280 & 69:02:07.570 & 1.82 & -111 & 13 \\
1308 & -- & 50175 & -- & 09:55:37.770 & 68:59:17.690 & 4.61 & -194 & 22 \\
1309 & -- & 50463 & -- & 09:55:37.800 & 69:03:27.960 & 0.61 & 53 & 25 \\
1327 & -- & 50802 & -- & 09:55:38.510 & 69:06:55.020 & 3.02 & 185 & 28 \\
1341 & -- & 50693 & -- & 09:55:39.350 & 69:05:32.620 & 1.71 & 162 & 47 \\
1350 & -- & 50387 & -- & 09:55:40.010 & 69:02:29.560 & 1.55 & -368 & 51 \\
1352 & -- & -- & -- & 09:55:40.020 & 69:04:10.140 & 0.66 & 353 & 12 \\
1363 & -- & 50681 & -- & 09:55:40.490 & 69:05:24.700 & 1.62 & 179 & 44 \\
1393 & BH91 HS26 & 60012 & -- & 09:55:41.930 & 68:55:00.740 & 8.90 & -150 & 17 \\
1413 & -- & 50512 & -- & 09:55:43.400 & 69:03:51.640 & 0.91 & -102 & 16 \\
1428 & -- & 50583 & -- & 09:55:44.230 & 69:04:24.130 & 1.09 & 101 & 30 \\
1456 & -- & 50315 & -- & 09:55:46.040 & 69:01:25.730 & 2.73 & -255 & 22 \\
1490 & -- & 50760 & -- & 09:55:47.680 & 69:06:25.230 & 2.80 & -41 & 18 \\
1495 & -- & 50858 & -- & 09:55:48.000 & 69:07:27.830 & 3.77 & 25 & 75 \\
1496 & SBKHP 07 & 50514 & -- & 09:55:48.000 & 69:03:52.020 & 1.31 & -218 & 33 \\
1506 & -- & 50744 & -- & 09:55:48.500 & 69:06:11.990 & 2.65 & 163 & 41 \\
1512 & -- & 50674 & 64 & 09:55:48.760 & 69:05:22.260 & 2.00 & -51 & 28 \\
1524 & SBKHP 02 & 50304 & -- & 09:55:49.180 & 69:01:15.350 & 3.00 & 50 & 14 \\
1537 & -- & 50111 & -- & 09:55:50.260 & 68:58:22.800 & 5.72 & -215 & 45 \\
1563 & SBKHP 06 & 50460 & -- & 09:55:51.320 & 69:03:23.650 & 1.69 & 207 & 26 \\
1571 & SBKHP 15 & 50889 & 7 & 09:55:51.860 & 69:07:39.490 & 4.07 & -61 & 12 \\
1627 & SBKHP 05 & 50418 & -- & 09:55:54.510 & 69:02:52.620 & 2.16 & -29 & 15 \\
1635 & SBKHP 01 & 50285 & -- & 09:55:54.980 & 69:00:56.160 & 3.54 & -203 & 14 \\
1643 & -- & 50486 & -- & 09:55:55.260 & 69:03:37.520 & 1.98 & -84 & 17 \\
1652 & -- & 50386 & -- & 09:55:56.190 & 69:02:28.500 & 2.50 & -289 & 58 \\
1816 & -- & 50847 & 5 & 09:56:03.060 & 69:07:19.550 & 4.30 & 11 & 19 \\
1859 & -- & 51046 & -- & 09:56:05.010 & 69:09:21.290 & 6.10 & 114 & 17 \\
1946 & -- & 50381 & -- & 09:56:08.690 & 69:02:24.530 & 3.50 & -199 & 35 \\
1951 & -- & 50258 & -- & 09:56:08.820 & 69:00:23.580 & 4.73 & -77 & 57 \\
2081 & -- & 50651 & 76 & 09:56:14.140 & 69:05:05.320 & 3.82 & -235 & 67 \\
2087 & -- & 50320 & -- & 09:56:14.320 & 69:01:29.950 & 4.38 & -168 & 58 \\
2163 & -- & 50065 & -- & 09:56:17.510 & 68:57:12.060 & 7.77 & -159 & 32 \\
2170 & -- & 50178 & -- & 09:56:17.770 & 68:59:18.680 & 6.07 & -3 & 22 \\
2171 & -- & -- & -- & 09:56:17.780 & 69:03:04.620 & 4.06 & -134 & 25 \\
2196 & -- & 50228 & -- & 09:56:18.950 & 68:59:55.330 & 5.70 & -93 & 29 \\
2219 & PBH 50415 & 50415 & -- & 09:56:20.490 & 69:02:49.050 & 4.35 & -346 & 31 \\
2230 & (PBH 50355) & 50355 & -- & 09:56:21.090 & 69:02:01.510 & 4.66 & 63 & 18 \\
2327 & -- & 50301 & -- & 09:56:27.470 & 69:01:09.900 & 5.56 & -42 & 12 \\
2355 & (PBH 50254) & 50254 & -- & 09:56:31.730 & 69:03:54.870 & 5.20 & -282 & 32 \\
2381 & -- & 50338 & -- & 09:56:36.890 & 69:01:46.350 & 6.06 & -102 & 15 \\
2490 & -- & 50058 & -- & 09:44:01.840 & 68:56:52.910 & 10.57 & -113 & 15 \\
30244 & lit & 30244 & -- & 09:44:54.850 & 68:49:00.400 & 19.55 & 100 & 70 \\
40083 & lit & 40083 & -- & 09:56:38.470 & 69:22:50.300 & 19.67 & -87 & 52 \\
40165 & lit & 40165 & -- & 09:55:03.840 & 69:15:37.800 & 11.93 & 6 & 30 \\
40181 & lit & 40181 & -- & 09:55:44.090 & 69:14:11.700 & 10.27 & 46 & 30 \\
50037 & lit & 50037 & -- & 09:55:06.360 & 68:56:26.000 & 7.82 & -18 & 66 \\
50225 & lit & 50225 & -- & 09:56:40.590 & 68:59:52.600 & 7.23 & -7 & 13 \\
50233 & lit & 50233 & -- & 09:55:55.830 & 69:00:03.400 & 4.34 & -14 & 48 \\
50286 & lit & 50286 & -- & 09:54:58.870 & 69:00:58.200 & 4.24 & -9 & 68 \\
50357 & lit & 50357 & -- & 09:54:11.220 & 69:02:06.600 & 7.51 & 122 & 45 \\
50394 & lit & 50394 & -- & 09:54:16.500 & 69:02:34.400 & 6.95 & -70 & 30 \\
50401 & lit & 50401 & -- & 09:56:31.750 & 69:02:38.400 & 5.36 & -283 & 87 \\
50696 & lit & 50696 & -- & 09:55:02.450 & 69:05:38.100 & 3.22 & -44 & 28 \\
50785 & lit & 50785 & -- & 09:54:35.880 & 69:06:43.100 & 5.79 & 197 & 48 \\
50826 & lit & 50826 & -- & 09:55:52.190 & 69:07:10.400 & 3.65 & -22 & 80 \\
50886 & lit & 50886 & -- & 09:55:30.990 & 69:07:38.600 & 3.71 & 134 & 39 \\
50960 & lit & 50960 & -- & 09:55:52.060 & 69:08:18.700 & 4.68 & 77 & 43 \\
51027 & lit & 51027 & -- & 09:54:20.020 & 69:09:11.100 & 8.33 & 300 & 97 \\
60045 & lit & 60045 & -- & 09:55:56.860 & 68:52:13.400 & 11.83 & -28 & 19 \\
70319 & lit & 70319 & -- & 09:53:42.910 & 69:13:23.900 & 13.55 & -242 & 68 \\
70349 & lit & 70349 & -- & 09:53:03.240 & 69:13:47.400 & 16.47 & -50 & 36 \\
80172 & lit & 80172 & -- & 09:53:51.690 & 68:57:04.400 & 11.34 & -51 & 40 \\
90103 & lit & 90103 & -- & 09:53:39.840 & 68:48:00.070 & 18.83 & -111 & 71 \\
BH91 HS01 & lit & 40312 & -- & 09:54:52.010 & 69:19:45.300 & 16.16 & 258 & 27 \\
BH91 HS02 & lit & 50864 & -- & 09:56:25.740 & 69:07:27.800 & 5.84 & 180 & 94 \\
BH91 HS06 & lit & 60072 & -- & 09:55:08.080 & 68:49:56.500 & 14.09 & 122 & 77 \\
BH91 R012 & lit & 60126 & -- & 09:55:26.970 & 68:44:53.600 & 18.93 & -37 & 39 \\
S04 & lit & 50378 & -- & 09:55:57.872 & 69:02:23.070 & 2.68 & -318 & 31 \\
S08 & lit & 50667 & -- & 09:55:22.295 & 69:05:19.160 & 1.70 & -144 & 9 \\
S09 & lit & 50690 & -- & 09:55:21.587 & 69:05:31.980 & 1.91 & -261 & 27 \\
S10 & lit & 50738 & -- & 09:55:30.277 & 69:06:06.120 & 2.19 & 22 & 39 \\
S11 & lit & 50759 & -- & 09:55:35.840 & 69:06:25.530 & 2.51 & 11 & 24 \\
S12 & lit & 50782 & -- & 09:55:33.042 & 69:06:39.880 & 2.73 & -200 & 17 \\
S13 & lit & 50787 & 1 & 09:56:05.569 & 69:06:42.970 & 4.00 & -87 & 49 \\
S16 & lit & 50889 & -- & 09:55:40.194 & 69:07:30.820 & 3.63 & -100 & 25 \\

\enddata


\tablecomments{Literature spectroscopy IDs in parentheses represent objects for which low signal-to-noise spectra were obtained by PBH, but the objects were not confirmed as clusters.  PR/PBH object 50388 was originally labeled as a star, but we find it to be a GC.  Object 1859, a GC overlapping with an H II region, was cleaned of emission lines before its velocity was determined, so as to find the velocity of the GC rather than the H II region.}

\end{deluxetable}

\begin{deluxetable}{ccccccccc}


\tabletypesize{\scriptsize}


\tablecaption{Non-GC Objects}


\tablehead{\colhead{ID} & \colhead{Obj. Type} & \colhead{PR95 ID} & \colhead{CFT01 ID} & \colhead{RA} & \colhead{Dec} & \colhead{Dist.} & \colhead{RV} & \colhead{$\sigma_{RV}$} \\ 
\colhead{} & \colhead{} & \colhead{} & \colhead{} & \colhead{(hrs.)} & \colhead{(deg.)} & \colhead{($\arcmin$)} & \colhead{(km s$^{-1}$)} & \colhead{(km s$^{-1}$)} } 

\startdata
36 & gx & -- & -- & 09 54 06.64 & 69 06 20.90 & 8.09 & 112001 & 28 \\
143 & hii & -- & -- & 09 54 34.65 & 69 05 54.56  & 5.59 & 85 & 10 \\
159 & oc & 50618 & --  & 09 54 36.76 & 69 04 46.96 & 5.11 & 76 & 59 \\
164 & hii & 50774 & -- & 09 54 37.67 & 69 06 36.90  & 5.64 & 103 & 14 \\
185 & hii & 50649 & --  & 09 54 39.63 & 69 05 07.22  & 4.93 & 71 & 11 \\
215 & hii & 50485 & --  & 09 54 42.47 & 69 03 38.25  & 4.54 & 32 & 10 \\
269 & hii & 50574 & --  & 09 54 46.63 & 69 04 22.32  & 4.18 & 33 & 9 \\
283 & hii & -- & -- & 09 54 47.58 & 69 03 23.27  & 4.11 & 13 & 10 \\
287 & hii & --     & --  & 09 54 47.72 & 69 03 22.52  & 4.10 & 5 & 14 \\
301 & hii & 50729 & --  & 09 54 48.53 & 69 05 59.46  & 4.49 & 114 & 12 \\
329 & hii & 50840 & --  & 09 54 49.30 & 69 07 18.78  & 5.18 & 139 & 12 \\
357 & hii & --     & --  & 09 54 50.98 & 69 02 50.89  & 3.92 & -20 & 10 \\
371 & hii & -- & -- & 09 54 51.60 & 69 06 35.39  & 4.57 & 106 & 12 \\
395 & hii & 50997 & --  & 09 54 52.97 & 69 08 50.38  & 6.08 & 173 & 13 \\
406 & hii & 50883 & --  & 09 54 53.42 & 69 07 38.63  & 5.14 & 154 & 13 \\
469 & hii & 50996 & --  & 09 54 56.85 & 69 08 49.26  & 5.87 & 165 & 11 \\
540 & hii & 50183 & --  & 09 55 00.69 & 68 59 22.77  & 5.39 & -103 & 15 \\
561 & hii & 40291 &  & 09 55 01.58 & 69 12 56.19  & 9.44 & 131 & 13 \\
611 & hii & 51117 & -- & 09 55 03.45 & 69 10 54.07  & 7.47 & 168 & 11 \\
629 & hii & 50659 & 41 & 09 55 04.43 & 69 09 56.14  & 6.54 & 139 & 24 \\
668 & gx & 51043 & 32 & 09 55 06.65 & 69 09 20.11 & 5.91 & 81798 & 51 \\
802 & gx & 50331 & --  & 09 55 12.75 & 69 01 41.81 & 2.88 & 69171 & 16 \\
811 & gx & --     & 103 & 09 55 13.25 & 69 01 34.41 & 2.95 & 61522 & 23 \\
859 & oc & 50979 &  & 09 55 15.02 & 69 08 31.42 & 4.88 & 187 & 46 \\
889 & hii & 51060 & -- & 09 55 16.17 & 69 09 35.11  & 5.87 & 168 & 21 \\
898 & hii & 51003 & --  & 09 55 16.52 & 69 08 55.17  & 5.22 & 165 & 11 \\
944 & hii & 50975 & --  & 09 55 18.40 & 69 08 28.75  & 4.75 & 139 & 11 \\
956 & hii & --     & --  & 09 55 18.92 & 69 10 05.76  & 6.31 & 156 & 14 \\
991 & hii & 50371 & 105 & 09 55 20.04 & 69 02 18.56  & 1.99 & -74 & 17 \\
1011 & hii & 50480 & --  & 09 55 21.17 & 69 03 36.32  & 1.12 & -20 & 17 \\
1027 & hii & --     & 54 & 09 55 21.89 & 69 07 36.52  & 3.83 & 139 & 14 \\
1036 & gx & 50589 & --  & 09 55 22.27 & 69 04 26.39 & 1.10 & 99145 & 18 \\
1121 & hii & 50577 & --  & 09 55 26.87 & 69 04 22.15  & 0.72 & 213 & 18 \\
1227 & hii & --     & --  & 09 55 32.76 & 69 11 55.63  & 8.01 & 101 & 12 \\
1252 & hii & 50980 & --  & 09 55 34.21 & 69 08 31.50  & 4.61 & 117 & 11 \\
1426 & hii & 50845 & --  & 09 55 44.17 & 69 07 19.27  & 3.54 & 64 & 13 \\
1470 & hii & 50711 & --  & 09 55 46.85 & 69 05 48.86  & 2.26 & 40 & 24 \\
1471 & gx & --     & --  & 09 55 46.86 & 68 55 43.02 & 8.29 & 124246 & 68 \\
1473 & hii & 50179 & --  & 09 55 46.88 & 68 59 20.18  & 4.74 & -223 & 12 \\
1479 & gx & --     & --  & 09 55 47.20 & 68 53 56.79 & 10.05 & 82813 & 71 \\
1497 & gx & 50775 & --  & 09 55 48.09 & 69 06 35.77 & 2.99 & 100229 & 79 \\
1515 & hii & 50113 & --  & 09 55 48.89 & 68 58 27.49  & 5.64 & -235 & 12 \\
1593 & hii & 50152 & --  & 09 55 52.98 & 68 59 04.18  & 5.16 & -234 & 14 \\
1673 & hii & 50633 & --  & 09 55 57.70 & 69 04 55.54  & 2.41 & -36 & 13 \\
1685 & hii & 51008 & --  & 09 55 58.05 & 69 08 56.62  & 5.49 & 23 & 11 \\
1704 & hii & --     & --  & 09 55 58.69 & 69 09 09.36  & 5.71 & 20 & 10 \\
1734 & gx & -- & -- & 09 56 00.12 & 68 57 13.25 & 7.12 & 56775 & 21 \\
1736 & gx & --     & --  & 09 56 00.13 & 69 05 33.40 & 2.91 & 100396 & 62 \\
1743 & hii & 50531 & --  & 09 56 00.60 & 69 04 00.31  & 2.45 & -70 & 10 \\
1747 & gx & -- & -- & 09 56 00.75 & 68 56 18.14 & 8.01 & 124037 & 69 \\
1757 & hii & --     & --  & 09 56 01.03 & 68 59 02.00  & 5.49 & -233 & 13 \\
1772 & hii & 50098 & --  & 09 56 01.58 & 68 58 03.84  & 6.39 & -257 & 16 \\
1830 & hii & --     & --  & 09 56 03.58 & 69 08 24.64  & 5.25 & 2 & 25 \\
1839 & hii & 51067 & --  & 09 56 04.01 & 69 09 39.84  & 6.37 & 29 & 63 \\
1843 & gx & --     & --  & 09 56 04.17 & 68 56 13.83 & 8.18 & 123902 & 42 \\
1889 & hii & 50200 & --  & 09 56 06.19 & 68 59 34.23  & 5.26 & -228 & 14 \\
1922 & oc & 50430 & -- & 09 56 08.11 & 69 03 05.31 & 3.23 & -81 & 46 \\
1934 & hii & --     & 75 & 09 56 08.28 & 69 05 17.98  & 3.42 & -23 & 10 \\
1970 & gx & 50962 & --  & 09 56 09.56 & 69 08 19.47 & 5.47 & 32332 & 17 \\
1976 & hii & 50319 & --  & 09 56 09.80 & 69 01 29.73  & 4.08 & -186 & 13 \\
2002 & oc & 50584 & --  & 09 56 10.66 & 69 04 23.58 & 3.38 & 63 & 69 \\
2114 & oc & 50377 & --  & 09 56 15.33 & 69 02 22.19 & 4.08 & -149 & 67 \\
2146 & hii & --     & --  & 09 56 16.70 & 69 05 39.23  & 4.25 & -72 & 11 \\
2240 & oc & 50822 & --  & 09 56 21.32 & 69 07 07.83 & 5.36 & 24 & 85 \\
2288 & hii & 50176 & --  & 09 56 24.38 & 68 59 15.82  & 6.54 & -228 & 12 \\
2290 & oc & 50866 & --  & 09 56 24.49 & 69 07 28.86 & 5.80 & 8 & 20 \\
2412 & gx & 50170 & --  & 09 56 42.89 & 68 59 11.67 & 7.83 & 56450 & 36 \\
2462 & gx & -- & -- & 09 56 53.48 & 69 03 11.55 & 7.21 & 128234 & 76 \\
2468 & gx & -- & -- & 09 56 55.16 & 68 58 57.20 & 8.87 & 60258 & 15 \\
2471 & gx & -- & -- & 09 56 55.37 & 68 57 59.10 & 9.47 & 36073 & 28 \\
2479 & gx & -- & -- & 09 56 57.53 & 68 58 10.81 & 9.50 & 127370 & 56 \\
2484 & gx & -- & -- & 09 56 59.07 & 69 01 28.74 & 8.07 & 127833 & 98 \\
2499 & gx & 50107 & --  & 09 57 04.67 & 68 58 13.24 & 9.99 & 83038 & 65 \\
2500 & gx & 20246 & --  & 09 57 04.72 & 69 01 31.67 & 8.53 & 63144 & 10 \\
2510 & gx & 50238 & --  & 09 57 09.62 & 69 00 02.11 & 9.47 & 26763 & 18 \\
2518 & gx & 50120 & --  & 09 57 13.18 & 68 58 28.51 & 10.49 & 104981 & 209 \\
-- & hii & -- & -- & 09 54 40.90 & 69 05 01.85  & 4.80 & 44 & 14 \\
\enddata




\end{deluxetable}

\begin{deluxetable}{ccccccccccccc}


\tabletypesize{\scriptsize}


\tablecaption{Brodie and Huchra Indices}


\tablehead{\colhead{ID} & \colhead{CNR/CN1} & \colhead{G4300} & \colhead{H$\beta$} & \colhead{MgH/Mg1} & \colhead{Mg2} & \colhead{Mgb} & \colhead{Fe5270} & \colhead{NaD/Na1} & \colhead{CNB} & \colhead{H\&K} & \colhead{MgG} & \colhead{$\delta$} \\ 
\colhead{} & \colhead{(mag)} & \colhead{(mag)} & \colhead{(mag)} & \colhead{(mag)} & \colhead{(mag)} & \colhead{(mag)} & \colhead{(mag)} & \colhead{(mag)} & \colhead{(mag)} & \colhead{(mag)} & \colhead{(mag)} & \colhead{(mag)} } 

\startdata
34 & 0.510 & 0.534 & 0.014 & 0.077 & 0.237 & 0.174 & 0.102 & 0.129 & nan & 1.287 & -0.424 & 2.954 \\
$\sigma$ & 0.169 & 0.227 & 0.052 & 0.029 & 0.033 & 0.058 & 0.042 & 0.060 & 1.598 & 0.667 & 0.533 & 0.525 \\
108 & 0.002 & 0.288 & 0.097 & 0.033 & 0.086 & 0.090 & 0.116 & 0.040 & 0.184 & 0.533 & -0.260 & 0.535 \\
$\sigma$ & 0.083 & 0.091 & 0.050 & 0.027 & 0.029 & 0.051 & 0.041 & 0.066 & 0.207 & 0.174 & 0.147 & 0.048 \\
173 & 0.116 & 0.249 & 0.066 & 0.125 & 0.266 & 0.179 & 0.075 & 0.149 & 0.097 & 0.289 & -0.232 & 0.563 \\
$\sigma$ & 0.060 & 0.068 & 0.030 & 0.017 & 0.019 & 0.033 & 0.025 & 0.037 & 0.177 & 0.148 & 0.111 & 0.035 \\
187 & -0.108 & 0.103 & 0.082 & 0.039 & 0.020 & 0.000 & -0.003 & 0.101 & 0.102 & 0.362 & -0.318 & 0.348 \\
$\sigma$ & 0.073 & 0.079 & 0.043 & 0.024 & 0.025 & 0.043 & 0.035 & 0.043 & 0.157 & 0.118 & 0.128 & 0.033 \\
264 & -0.118 & 0.050 & 0.099 & 0.011 & 0.030 & 0.057 & 0.076 & 0.097 & -0.134 & 0.130 & -0.335 & 0.247 \\
$\sigma$ & 0.078 & 0.089 & 0.056 & 0.031 & 0.032 & 0.056 & 0.046 & 0.084 & 0.217 & 0.145 & 0.184 & 0.037 \\
345 & 0.006 & 0.112 & 0.063 & -0.065 & -0.081 & -0.089 & 0.008 & 0.132 & -0.031 & 0.225 & -0.395 & 0.177 \\
$\sigma$ & 0.173 & 0.197 & 0.124 & 0.068 & 0.071 & 0.128 & 0.105 & 0.160 & 0.384 & 0.301 & 0.307 & 0.078 \\
359 & -0.093 & 0.352 & 0.075 & 0.077 & 0.208 & 0.199 & 0.053 & 0.101 & 0.437 & 0.598 & -0.322 & 0.667 \\
$\sigma$ & 0.113 & 0.122 & 0.053 & 0.028 & 0.031 & 0.053 & 0.042 & 0.055 & 0.293 & 0.239 & 0.200 & 0.073 \\
464 & -0.056 & 0.125 & 0.018 & 0.098 & 0.163 & 0.110 & 0.124 & 0.076 & 0.047 & 0.297 & -0.346 & 0.438 \\
$\sigma$ & 0.089 & 0.111 & 0.052 & 0.030 & 0.032 & 0.056 & 0.044 & 0.066 & 0.276 & 0.191 & 0.229 & 0.050 \\
505 & 0.066 & 0.075 & 0.252 & 0.110 & 0.160 & 0.120 & 0.007 & 0.156 & 0.019 & 0.137 & -0.272 & 0.582 \\
$\sigma$ & 0.120 & 0.144 & 0.082 & 0.041 & 0.045 & 0.079 & 0.064 & 0.090 & 0.398 & 0.342 & 0.262 & 0.078 \\
526 & 0.633 & 0.147 & 0.132 & 0.098 & -0.028 & 0.274 & 0.149 & 0.104 & 0.593 & 0.310 & -0.287 & 0.181 \\
$\sigma$ & 0.543 & 0.437 & 0.195 & 0.122 & 0.121 & 0.195 & 0.165 & 0.208 & 0.565 & 4.372 & 0.473 & 0.191 \\
594 & 0.054 & 0.035 & 0.059 & 0.128 & 0.105 & -0.015 & 0.134 & 0.096 & -0.947 & 0.302 & -0.930 & 0.516 \\
$\sigma$ & 0.158 & 0.164 & 0.082 & 0.043 & 0.046 & 0.084 & 0.057 & 0.058 & 0.672 & 0.300 & 0.626 & 0.083 \\
605 & -0.197 & 0.174 & 0.079 & 0.038 & 0.035 & -0.053 & 0.060 & -0.096 & -0.397 & -0.119 & -0.242 & 0.164 \\
$\sigma$ & 0.115 & 0.132 & 0.084 & 0.045 & 0.048 & 0.087 & 0.068 & 0.107 & 0.297 & 0.261 & 0.192 & 0.049 \\
628 & -0.075 & 0.025 & 0.044 & 0.033 & 0.088 & 0.044 & 0.090 & 0.112 & 0.034 & 0.424 & -0.301 & 0.465 \\
$\sigma$ & 0.054 & 0.070 & 0.033 & 0.019 & 0.021 & 0.037 & 0.029 & 0.033 & 0.126 & 0.107 & 0.090 & 0.028 \\
676 & 0.789 & 0.012 & 0.246 & 0.185 & 0.238 & 0.368 & 0.099 & 0.103 & 1.329 & nan & -0.286 & 2.248 \\
$\sigma$ & 0.576 & 0.456 & 0.145 & 0.073 & 0.080 & 0.132 & 0.112 & 0.145 & 2.750 & 1.694 & 2.335 & 2.096 \\
705 & -0.008 & 0.116 & 0.020 & 0.064 & 0.127 & 0.078 & 0.064 & 0.086 & 0.079 & 0.328 & -0.209 & 0.471 \\
$\sigma$ & 0.076 & 0.096 & 0.049 & 0.029 & 0.031 & 0.054 & 0.045 & 0.060 & 0.182 & 0.164 & 0.116 & 0.040 \\
720 & -0.091 & 0.148 & 0.040 & 0.061 & 0.107 & -0.007 & 0.041 & 0.218 & 0.393 & 0.556 & -0.487 & 0.657 \\
$\sigma$ & 0.104 & 0.100 & 0.048 & 0.025 & 0.027 & 0.047 & 0.035 & 0.032 & 0.284 & 0.181 & 0.242 & 0.057 \\
722 & -0.059 & 0.069 & 0.088 & 0.008 & 0.018 & 0.054 & 0.062 & 0.092 & 0.216 & 0.306 & -0.218 & 0.398 \\
$\sigma$ & 0.037 & 0.044 & 0.025 & 0.014 & 0.015 & 0.025 & 0.021 & 0.024 & 0.075 & 0.068 & 0.052 & 0.018 \\
743 & 0.043 & 0.106 & 0.103 & 0.036 & 0.147 & 0.122 & 0.076 & 0.175 & 0.191 & 0.342 & -0.228 & 0.448 \\
$\sigma$ & 0.012 & 0.014 & 0.008 & 0.005 & 0.005 & 0.009 & 0.007 & 0.008 & 0.026 & 0.022 & 0.020 & 0.006 \\
839 & -0.033 & 0.022 & 0.078 & 0.002 & 0.051 & 0.121 & 0.026 & 0.081 & 0.046 & 0.219 & -0.196 & 0.317 \\
$\sigma$ & 0.053 & 0.067 & 0.038 & 0.022 & 0.024 & 0.039 & 0.035 & 0.049 & 0.123 & 0.102 & 0.089 & 0.026 \\
861 & 0.113 & 0.145 & 0.089 & 0.049 & 0.165 & 0.154 & 0.057 & 0.091 & 0.053 & 0.186 & -0.303 & 0.544 \\
$\sigma$ & 0.061 & 0.074 & 0.033 & 0.019 & 0.021 & 0.036 & 0.029 & 0.041 & 0.192 & 0.167 & 0.122 & 0.037 \\
863 & 0.066 & 0.141 & 0.060 & 0.022 & 0.131 & 0.122 & 0.085 & 0.111 & 0.045 & 0.433 & -0.266 & 0.463 \\
$\sigma$ & 0.108 & 0.130 & 0.062 & 0.035 & 0.038 & 0.064 & 0.053 & 0.060 & 0.249 & 0.208 & 0.179 & 0.056 \\
876 & -0.034 & 0.350 & 0.098 & 0.139 & 0.127 & 0.073 & -0.002 & 0.002 & -0.187 & 0.688 & -0.524 & 0.436 \\
$\sigma$ & 0.133 & 0.157 & 0.091 & 0.053 & 0.056 & 0.098 & 0.079 & 0.080 & 0.561 & 0.224 & 0.534 & 0.069 \\
966 & 0.027 & 0.129 & -0.018 & 0.098 & 0.191 & 0.234 & 0.045 & 0.154 & 0.350 & 0.336 & -0.326 & 0.566 \\
$\sigma$ & 0.129 & 0.150 & 0.070 & 0.040 & 0.044 & 0.075 & 0.062 & 0.063 & 0.351 & 0.254 & 0.300 & 0.071 \\
993 & 0.093 & 0.109 & 0.127 & 0.075 & 0.041 & 0.089 & 0.070 & 0.082 & 0.289 & 0.331 & -0.128 & 0.355 \\
$\sigma$ & 0.083 & 0.099 & 0.058 & 0.035 & 0.036 & 0.059 & 0.051 & 0.064 & 0.157 & 0.139 & 0.106 & 0.039 \\
1029 & 0.044 & 0.138 & 0.077 & 0.034 & 0.114 & 0.111 & 0.076 & 0.106 & 0.189 & 0.314 & -0.230 & 0.457 \\
$\sigma$ & 0.007 & 0.008 & 0.005 & 0.003 & 0.003 & 0.005 & 0.004 & 0.004 & 0.014 & 0.012 & 0.010 & 0.003 \\
1089 & -0.053 & 0.042 & 0.083 & 0.005 & 0.084 & 0.060 & 0.069 & 0.043 & 0.157 & 0.300 & -0.072 & 0.357 \\
$\sigma$ & 0.065 & 0.078 & 0.045 & 0.026 & 0.028 & 0.049 & 0.040 & 0.049 & 0.133 & 0.126 & 0.082 & 0.031 \\
1104 & -0.281 & -0.214 & 0.186 & 0.057 & 0.045 & 0.086 & 0.140 & 0.139 & 0.344 & -0.045 & -0.318 & 0.548 \\
$\sigma$ & 0.208 & 0.306 & 0.101 & 0.052 & 0.055 & 0.093 & 0.078 & 0.085 & 0.537 & 0.842 & 0.287 & 0.134 \\
1154 & -0.003 & 0.143 & 0.183 & 0.107 & 0.211 & 0.088 & 0.197 & 0.026 & -0.230 & 0.568 & -0.373 & 0.595 \\
$\sigma$ & 0.171 & 0.226 & 0.097 & 0.053 & 0.059 & 0.108 & 0.084 & 0.112 & 0.646 & 0.509 & 0.423 & 0.105 \\
1162 & 0.037 & 0.112 & 0.070 & 0.027 & 0.111 & 0.133 & 0.072 & 0.106 & 0.218 & 0.414 & -0.318 & 0.567 \\
$\sigma$ & 0.033 & 0.042 & 0.022 & 0.013 & 0.014 & 0.023 & 0.020 & 0.021 & 0.078 & 0.065 & 0.057 & 0.018 \\
1172 & 0.030 & 0.214 & -0.024 & -0.050 & 0.051 & 0.070 & 0.064 & 0.093 & 0.257 & 0.295 & -0.318 & 0.522 \\
$\sigma$ & 0.108 & 0.133 & 0.076 & 0.043 & 0.047 & 0.077 & 0.062 & 0.079 & 0.235 & 0.233 & 0.156 & 0.058 \\
1257 & -0.088 & -0.164 & 0.124 & 0.004 & 0.005 & 0.021 & -0.021 & 0.144 & -0.095 & 0.047 & -0.346 & 0.169 \\
$\sigma$ & 0.164 & 0.241 & 0.108 & 0.057 & 0.061 & 0.103 & 0.093 & 0.102 & 0.369 & 0.366 & 0.255 & 0.081 \\
1265 & 0.039 & 0.172 & 0.073 & 0.024 & 0.116 & 0.139 & 0.060 & 0.129 & 0.217 & 0.284 & -0.224 & 0.548 \\
$\sigma$ & 0.048 & 0.058 & 0.028 & 0.016 & 0.017 & 0.029 & 0.024 & 0.028 & 0.122 & 0.109 & 0.077 & 0.027 \\
1300 & 0.017 & 0.081 & 0.048 & -0.030 & -0.048 & 0.040 & 0.023 & 0.091 & 0.023 & 0.254 & -0.281 & 0.219 \\
$\sigma$ & 0.079 & 0.084 & 0.046 & 0.024 & 0.025 & 0.040 & 0.035 & 0.033 & 0.145 & 0.138 & 0.102 & 0.035 \\
1301 & 0.107 & 0.191 & 0.076 & 0.071 & 0.181 & 0.159 & 0.070 & 0.208 & 0.363 & 0.553 & -0.263 & 0.746 \\
$\sigma$ & 0.041 & 0.047 & 0.024 & 0.014 & 0.015 & 0.026 & 0.021 & 0.021 & 0.108 & 0.092 & 0.072 & 0.025 \\
1308 & 0.112 & 0.253 & 0.027 & 0.043 & 0.145 & 0.185 & 0.085 & 0.051 & 0.157 & 0.310 & -0.259 & 0.625 \\
$\sigma$ & 0.075 & 0.091 & 0.041 & 0.024 & 0.026 & 0.044 & 0.035 & 0.051 & 0.233 & 0.217 & 0.125 & 0.047 \\
1309 & 0.001 & 0.119 & 0.074 & 0.003 & 0.025 & 0.011 & 0.046 & 0.108 & 0.150 & 0.273 & -0.158 & 0.293 \\
$\sigma$ & 0.026 & 0.032 & 0.023 & 0.013 & 0.014 & 0.023 & 0.021 & 0.022 & 0.044 & 0.042 & 0.033 & 0.011 \\
1327 & 0.002 & 0.016 & 0.105 & 0.019 & 0.043 & 0.031 & 0.110 & 0.081 & -0.036 & 0.297 & -0.437 & 0.625 \\
$\sigma$ & 0.085 & 0.098 & 0.039 & 0.021 & 0.022 & 0.038 & 0.031 & 0.035 & 0.255 & 0.209 & 0.170 & 0.049 \\
1341 & 0.005 & 0.069 & 0.170 & 0.029 & 0.002 & 0.058 & 0.091 & 0.046 & 0.217 & 0.248 & -0.124 & 0.195 \\
$\sigma$ & 0.087 & 0.097 & 0.066 & 0.037 & 0.038 & 0.064 & 0.056 & 0.062 & 0.139 & 0.134 & 0.105 & 0.037 \\
1350 & 0.080 & 0.063 & 0.066 & 0.030 & 0.079 & -0.025 & 0.051 & -0.001 & -0.026 & 0.296 & -0.585 & 0.530 \\
$\sigma$ & 0.164 & 0.241 & 0.125 & 0.080 & 0.086 & 0.159 & 0.130 & 0.142 & 1.067 & 0.322 & 1.038 & 0.088 \\
1352 & -0.059 & 0.060 & 0.068 & -0.025 & 0.017 & 0.050 & 0.030 & 0.053 & 0.041 & 0.280 & -0.171 & 0.267 \\
$\sigma$ & 0.014 & 0.018 & 0.012 & 0.007 & 0.008 & 0.012 & 0.011 & 0.012 & 0.025 & 0.023 & 0.020 & 0.006 \\
1363 & 0.030 & 0.091 & 0.130 & 0.017 & -0.024 & -0.020 & 0.015 & 0.020 & 0.102 & 0.193 & -0.304 & 0.433 \\
$\sigma$ & 0.110 & 0.127 & 0.066 & 0.034 & 0.036 & 0.058 & 0.048 & 0.050 & 0.252 & 0.259 & 0.128 & 0.058 \\
1393 & 0.065 & 0.211 & 0.092 & 0.102 & 0.213 & 0.128 & 0.088 & 0.143 & 0.277 & 0.332 & -0.267 & 0.620 \\
$\sigma$ & 0.061 & 0.073 & 0.032 & 0.018 & 0.020 & 0.035 & 0.027 & 0.042 & 0.174 & 0.141 & 0.123 & 0.038 \\
1413 & 0.059 & 0.126 & 0.028 & 0.035 & 0.170 & 0.177 & 0.083 & 0.124 & 0.288 & 0.527 & -0.356 & 0.746 \\
$\sigma$ & 0.055 & 0.066 & 0.034 & 0.019 & 0.021 & 0.035 & 0.029 & 0.030 & 0.154 & 0.110 & 0.120 & 0.032 \\
1428 & -0.007 & 0.154 & 0.140 & 0.056 & 0.105 & 0.128 & 0.029 & 0.086 & 0.313 & 0.287 & -0.363 & 0.389 \\
$\sigma$ & 0.115 & 0.136 & 0.068 & 0.037 & 0.039 & 0.066 & 0.054 & 0.061 & 0.239 & 0.195 & 0.178 & 0.056 \\
1456 & 0.142 & 0.192 & 0.006 & 0.022 & 0.129 & 0.145 & 0.035 & 0.088 & 0.186 & 0.427 & -0.307 & 0.553 \\
$\sigma$ & 0.070 & 0.080 & 0.043 & 0.023 & 0.026 & 0.043 & 0.037 & 0.045 & 0.166 & 0.140 & 0.118 & 0.038 \\
1490 & -0.067 & 0.131 & 0.078 & 0.012 & 0.027 & 0.051 & 0.025 & 0.055 & 0.007 & 0.259 & -0.228 & 0.374 \\
$\sigma$ & 0.031 & 0.037 & 0.022 & 0.012 & 0.013 & 0.023 & 0.019 & 0.022 & 0.068 & 0.060 & 0.049 & 0.015 \\
1495 & -0.231 & 0.410 & -0.156 & 0.053 & 0.070 & 0.227 & -0.041 & 0.291 & 0.186 & 0.394 & -0.165 & 0.464 \\
$\sigma$ & 0.332 & 0.310 & 0.180 & 0.101 & 0.107 & 0.170 & 0.167 & 0.163 & 0.640 & 0.859 & 0.420 & 0.173 \\
1496 & -0.093 & 0.103 & 0.095 & 0.093 & 0.184 & 0.046 & 0.025 & 0.158 & 0.467 & 0.603 & -0.353 & 0.846 \\
$\sigma$ & 0.108 & 0.130 & 0.063 & 0.034 & 0.037 & 0.071 & 0.054 & 0.053 & 0.300 & 0.242 & 0.195 & 0.066 \\
1506 & -0.365 & 0.270 & 0.188 & 0.233 & 0.224 & -0.021 & 0.081 & 0.360 & nan & 0.510 & -0.821 & nan \\
$\sigma$ & 0.397 & 0.376 & 0.124 & 0.069 & 0.074 & 0.150 & 0.095 & 0.128 & 11.712 & 11.149 & 3.636 & 0.559 \\
1512 & -0.094 & 0.024 & 0.113 & -0.005 & 0.029 & 0.052 & 0.031 & 0.068 & 0.110 & 0.144 & -0.108 & 0.260 \\
$\sigma$ & 0.045 & 0.058 & 0.036 & 0.020 & 0.022 & 0.037 & 0.032 & 0.044 & 0.089 & 0.083 & 0.060 & 0.021 \\
1524 & -0.049 & 0.146 & 0.072 & 0.021 & 0.066 & 0.092 & 0.051 & 0.120 & 0.109 & 0.399 & -0.307 & 0.530 \\
$\sigma$ & 0.039 & 0.047 & 0.024 & 0.014 & 0.015 & 0.025 & 0.021 & 0.023 & 0.092 & 0.078 & 0.066 & 0.021 \\
1537 & -0.047 & 0.040 & 0.021 & -0.021 & 0.016 & 0.085 & 0.057 & 0.077 & 0.054 & 0.256 & -0.052 & 0.144 \\
$\sigma$ & 0.052 & 0.062 & 0.038 & 0.022 & 0.023 & 0.038 & 0.033 & 0.041 & 0.094 & 0.082 & 0.072 & 0.022 \\
1563 & -0.024 & 0.053 & 0.090 & -0.002 & 0.082 & 0.056 & 0.074 & 0.034 & 0.129 & 0.288 & -0.072 & 0.255 \\
$\sigma$ & 0.052 & 0.065 & 0.039 & 0.022 & 0.024 & 0.041 & 0.034 & 0.037 & 0.090 & 0.088 & 0.067 & 0.023 \\
1571 & 0.043 & 0.123 & 0.056 & 0.041 & 0.144 & 0.164 & 0.072 & 0.150 & 0.322 & 0.346 & -0.294 & 0.663 \\
$\sigma$ & 0.035 & 0.041 & 0.021 & 0.012 & 0.013 & 0.023 & 0.018 & 0.023 & 0.098 & 0.079 & 0.072 & 0.022 \\
1627 & 0.017 & 0.163 & 0.065 & 0.035 & 0.082 & 0.076 & 0.071 & 0.027 & 0.180 & 0.294 & -0.256 & 0.529 \\
$\sigma$ & 0.027 & 0.032 & 0.017 & 0.010 & 0.011 & 0.019 & 0.015 & 0.016 & 0.063 & 0.056 & 0.041 & 0.014 \\
1635 & 0.091 & 0.117 & 0.055 & 0.011 & 0.115 & 0.133 & 0.050 & 0.129 & 0.204 & 0.442 & -0.267 & 0.487 \\
$\sigma$ & 0.044 & 0.052 & 0.025 & 0.014 & 0.016 & 0.027 & 0.021 & 0.023 & 0.102 & 0.084 & 0.075 & 0.023 \\
1643 & 0.169 & 0.324 & 0.072 & 0.088 & 0.214 & 0.148 & 0.134 & 0.135 & 0.585 & 0.200 & -0.277 & 0.654 \\
$\sigma$ & 0.082 & 0.095 & 0.043 & 0.022 & 0.025 & 0.043 & 0.033 & 0.039 & 0.229 & 0.199 & 0.123 & 0.050 \\
1652 & -0.067 & 0.131 & 0.114 & -0.024 & 0.006 & -0.022 & -0.023 & -0.062 & -0.152 & 0.364 & -0.230 & 0.242 \\
$\sigma$ & 0.098 & 0.116 & 0.075 & 0.045 & 0.048 & 0.082 & 0.072 & 0.090 & 0.215 & 0.180 & 0.164 & 0.045 \\
1816 & 0.129 & 0.170 & 0.066 & 0.050 & 0.153 & 0.141 & 0.080 & 0.086 & 0.407 & 0.367 & -0.290 & 0.642 \\
$\sigma$ & 0.070 & 0.079 & 0.036 & 0.019 & 0.021 & 0.036 & 0.029 & 0.039 & 0.189 & 0.161 & 0.123 & 0.044 \\
1859 & 0.025 & 0.146 & -0.083 & 0.067 & 0.149 & 0.129 & 0.065 & 0.132 & 0.113 & 0.389 & -0.277 & 0.527 \\
$\sigma$ & 0.045 & 0.055 & 0.024 & 0.014 & 0.016 & 0.027 & 0.022 & 0.032 & 0.124 & 0.102 & 0.084 & 0.026 \\
1946 & 0.007 & 0.180 & 0.263 & 0.026 & 0.085 & 0.051 & 0.077 & 0.087 & 0.491 & 0.496 & -0.112 & 0.325 \\
$\sigma$ & 0.086 & 0.101 & 0.058 & 0.031 & 0.033 & 0.056 & 0.045 & 0.061 & 0.153 & 0.145 & 0.106 & 0.044 \\
1951 & -0.099 & 0.356 & 0.040 & 0.018 & -0.037 & -0.048 & -0.083 & 0.527 & -0.245 & -0.013 & -0.482 & 0.189 \\
$\sigma$ & 0.196 & 0.245 & 0.146 & 0.076 & 0.079 & 0.136 & 0.114 & 0.143 & 0.544 & 0.458 & 0.483 & 0.087 \\
2081 & 0.199 & 0.056 & 0.117 & -0.015 & 0.121 & 0.102 & 0.033 & -0.019 & 0.118 & 0.238 & -0.076 & 0.186 \\
$\sigma$ & 0.102 & 0.120 & 0.068 & 0.037 & 0.041 & 0.070 & 0.058 & 0.084 & 0.195 & 0.169 & 0.140 & 0.045 \\
2087 & -0.102 & 0.146 & 0.145 & 0.022 & 0.060 & 0.022 & 0.030 & 0.068 & 0.240 & 0.351 & -0.106 & 0.253 \\
$\sigma$ & 0.058 & 0.069 & 0.041 & 0.023 & 0.025 & 0.042 & 0.036 & 0.054 & 0.102 & 0.092 & 0.076 & 0.026 \\
2163 & 0.103 & 0.161 & 0.084 & -0.009 & 0.009 & 0.044 & 0.090 & 0.026 & 0.157 & 0.338 & -0.325 & 0.381 \\
$\sigma$ & 0.071 & 0.077 & 0.042 & 0.023 & 0.025 & 0.043 & 0.034 & 0.044 & 0.156 & 0.121 & 0.124 & 0.034 \\
2170 & 0.012 & 0.120 & 0.099 & 0.005 & 0.100 & 0.120 & 0.040 & 0.052 & 0.120 & 0.410 & -0.302 & 0.485 \\
$\sigma$ & 0.050 & 0.063 & 0.032 & 0.018 & 0.019 & 0.033 & 0.027 & 0.038 & 0.130 & 0.103 & 0.101 & 0.028 \\
2171 & -0.033 & 0.138 & 0.140 & -0.003 & 0.044 & 0.016 & 0.051 & 0.066 & 0.184 & 0.324 & -0.278 & 0.367 \\
$\sigma$ & 0.052 & 0.059 & 0.032 & 0.018 & 0.019 & 0.034 & 0.028 & 0.040 & 0.112 & 0.097 & 0.082 & 0.026 \\
2196 & -0.183 & -0.013 & 0.113 & -0.011 & 0.025 & 0.016 & 0.050 & -0.011 & 0.146 & 0.368 & -0.147 & 0.297 \\
$\sigma$ & 0.053 & 0.065 & 0.037 & 0.021 & 0.022 & 0.038 & 0.032 & 0.049 & 0.102 & 0.090 & 0.076 & 0.025 \\
2219 & -0.018 & 0.079 & 0.076 & -0.020 & -0.008 & 0.020 & 0.015 & 0.076 & 0.013 & 0.396 & -0.215 & 0.288 \\
$\sigma$ & 0.038 & 0.047 & 0.029 & 0.016 & 0.017 & 0.030 & 0.025 & 0.030 & 0.082 & 0.066 & 0.064 & 0.018 \\
2230 & -0.029 & 0.099 & 0.067 & -0.008 & 0.035 & 0.048 & 0.043 & 0.023 & 0.100 & 0.300 & -0.147 & 0.297 \\
$\sigma$ & 0.026 & 0.033 & 0.019 & 0.011 & 0.012 & 0.021 & 0.017 & 0.020 & 0.051 & 0.046 & 0.037 & 0.012 \\
2327 & -0.059 & 0.115 & 0.079 & 0.001 & 0.033 & 0.029 & 0.039 & 0.068 & 0.067 & 0.265 & -0.076 & 0.256 \\
$\sigma$ & 0.016 & 0.019 & 0.012 & 0.007 & 0.008 & 0.013 & 0.011 & 0.012 & 0.030 & 0.028 & 0.021 & 0.007 \\
2355 & 0.135 & 0.290 & 0.024 & 0.067 & 0.178 & 0.163 & 0.107 & 0.014 & 0.138 & 0.313 & -0.136 & 0.526 \\
$\sigma$ & 0.073 & 0.087 & 0.044 & 0.025 & 0.028 & 0.047 & 0.036 & 0.060 & 0.208 & 0.186 & 0.122 & 0.043 \\
2381 & 0.128 & 0.189 & 0.066 & 0.066 & 0.195 & 0.124 & 0.067 & 0.106 & 0.074 & 0.292 & -0.336 & 0.579 \\
$\sigma$ & 0.044 & 0.050 & 0.023 & 0.013 & 0.014 & 0.026 & 0.020 & 0.029 & 0.142 & 0.110 & 0.097 & 0.026 \\
2490 & 0.020 & 0.163 & 0.082 & 0.035 & 0.081 & 0.057 & 0.064 & 0.105 & 0.149 & 0.223 & -0.268 & 0.443 \\
$\sigma$ & 0.035 & 0.040 & 0.021 & 0.012 & 0.012 & 0.021 & 0.017 & 0.027 & 0.089 & 0.078 & 0.057 & 0.019 \\
\enddata




\end{deluxetable}
\begin{deluxetable}{cccccccccccccc}


\tabletypesize{\scriptsize}


\tablecaption{Pseudo-Lick Indices}


\tablehead{\colhead{ID} & \colhead{CN2} & \colhead{Ca42} & \colhead{Fe43} & \colhead{Ca44} & \colhead{Fe45} & \colhead{C$_2$46} & \colhead{Fe50} & \colhead{Fe53} & \colhead{Fe54} & \colhead{Fe5709} & \colhead{Fe5782} & \colhead{TiO1} & \colhead{TiO2} \\ 
\colhead{} & \colhead{(mag)} & \colhead{(mag)} & \colhead{(mag)} & \colhead{(mag)} & \colhead{(mag)} & \colhead{(mag)} & \colhead{(mag)} & \colhead{(mag)} & \colhead{(mag)} & \colhead{(mag)} & \colhead{(mag)} & \colhead{(mag)} & \colhead{(mag)} } 

\startdata
34 & 0.485 & 0.577 & 0.124 & 0.203 & -0.013 & 0.036 & 0.082 & 0.074 & 0.136 & 0.081 & 0.075 & 0.060 & 0.061 \\
$\sigma$ & 0.206 & 0.333 & 0.156 & 0.118 & 0.087 & 0.063 & 0.052 & 0.051 & 0.052 & 0.049 & 0.058 & 0.030 & 0.022 \\
108 & 0.017 & 0.089 & 0.064 & 0.104 & 0.084 & 0.050 & 0.093 & 0.089 & 0.045 & 0.058 & 0.095 & 0.030 & 0.024 \\
$\sigma$ & 0.110 & 0.141 & 0.105 & 0.086 & 0.070 & 0.055 & 0.047 & 0.049 & 0.053 & 0.057 & 0.066 & 0.034 & 0.024 \\
173 & 0.155 & 0.150 & 0.121 & 0.125 & 0.108 & 0.128 & 0.084 & 0.128 & 0.081 & 0.013 & 0.091 & 0.031 & 0.084 \\
$\sigma$ & 0.080 & 0.096 & 0.064 & 0.053 & 0.043 & 0.033 & 0.029 & 0.029 & 0.031 & 0.031 & 0.035 & 0.019 & 0.014 \\
187 & -0.068 & -0.226 & 0.209 & 0.019 & 0.002 & -0.022 & 0.030 & 0.022 & 0.011 & 0.047 & 0.009 & 0.009 & 0.009 \\
$\sigma$ & 0.099 & 0.134 & 0.077 & 0.073 & 0.072 & 0.052 & 0.040 & 0.043 & 0.045 & 0.038 & 0.044 & 0.023 & 0.019 \\
264 & -0.075 & 0.176 & 0.035 & 0.108 & 0.006 & -0.008 & 0.080 & 0.091 & 0.009 & 0.037 & -0.055 & -0.027 & 0.020 \\
$\sigma$ & 0.102 & 0.144 & 0.105 & 0.085 & 0.075 & 0.061 & 0.052 & 0.059 & 0.064 & 0.068 & 0.080 & 0.043 & 0.033 \\
345 & 0.024 & 0.606 & 0.056 & -0.116 & -0.032 & -0.127 & 0.084 & 0.058 & -0.025 & 0.035 & 0.022 & 0.064 & -0.079 \\
$\sigma$ & 0.235 & 0.393 & 0.265 & 0.210 & 0.176 & 0.135 & 0.122 & 0.129 & 0.138 & 0.144 & 0.150 & 0.083 & 0.060 \\
359 & -0.058 & 0.374 & 0.147 & 0.049 & 0.031 & 0.164 & 0.139 & 0.063 & 0.015 & -0.017 & 0.067 & 0.019 & 0.047 \\
$\sigma$ & 0.151 & 0.189 & 0.107 & 0.094 & 0.076 & 0.054 & 0.048 & 0.049 & 0.052 & 0.052 & 0.055 & 0.030 & 0.022 \\
464 & -0.020 & 0.323 & -0.005 & 0.047 & 0.136 & 0.034 & 0.100 & 0.130 & 0.023 & 0.059 & 0.061 & 0.015 & 0.083 \\
$\sigma$ & 0.119 & 0.157 & 0.133 & 0.092 & 0.072 & 0.059 & 0.048 & 0.054 & 0.058 & 0.059 & 0.064 & 0.036 & 0.026 \\
505 & 0.207 & 0.052 & 0.133 & 0.085 & 0.071 & 0.057 & 0.049 & 0.085 & 0.051 & 0.014 & -0.039 & 0.011 & 0.016 \\
$\sigma$ & 0.141 & 0.244 & 0.131 & 0.124 & 0.110 & 0.082 & 0.077 & 0.078 & 0.084 & 0.083 & 0.094 & 0.049 & 0.036 \\
526 & 0.297 & -0.172 & 0.179 & 0.089 & 0.253 & -0.162 & 0.104 & -0.070 & -0.144 & -0.068 & 0.126 & -0.024 & 0.118 \\
$\sigma$ & 0.768 & 0.697 & 0.432 & 0.453 & 0.237 & 0.280 & 0.197 & 0.223 & 0.230 & 0.196 & 0.214 & 0.125 & 0.096 \\
594 & 0.260 & -0.092 & -0.380 & -0.170 & 0.080 & 0.036 & 0.252 & 0.082 & 0.094 & -0.059 & -0.019 & 0.102 & 0.051 \\
$\sigma$ & 0.184 & 0.245 & 0.220 & 0.178 & 0.113 & 0.095 & 0.076 & 0.070 & 0.065 & 0.055 & 0.063 & 0.033 & 0.024 \\
605 & -0.169 & -0.189 & -0.074 & 0.063 & 0.104 & 0.079 & -0.008 & 0.063 & 0.010 & -0.025 & 0.040 & -0.002 & -0.028 \\
$\sigma$ & 0.156 & 0.211 & 0.174 & 0.126 & 0.100 & 0.083 & 0.079 & 0.088 & 0.092 & 0.096 & 0.110 & 0.059 & 0.045 \\
628 & -0.053 & 0.065 & 0.037 & 0.030 & 0.052 & 0.051 & 0.071 & 0.008 & 0.005 & 0.036 & 0.015 & -0.007 & 0.030 \\
$\sigma$ & 0.074 & 0.097 & 0.064 & 0.063 & 0.052 & 0.039 & 0.032 & 0.037 & 0.037 & 0.031 & 0.038 & 0.019 & 0.016 \\
676 & 1.075 & 0.026 & -0.078 & -0.257 & 0.225 & 0.162 & 0.112 & 0.012 & -0.075 & 0.021 & 0.061 & 0.051 & 0.002 \\
$\sigma$ & 0.528 & 1.723 & 0.353 & 0.298 & 0.176 & 0.162 & 0.135 & 0.123 & 0.136 & 0.126 & 0.130 & 0.070 & 0.052 \\
705 & 0.003 & 0.241 & 0.027 & 0.129 & 0.094 & -0.002 & 0.062 & 0.052 & 0.057 & 0.012 & 0.103 & 0.011 & 0.051 \\
$\sigma$ & 0.104 & 0.123 & 0.105 & 0.084 & 0.069 & 0.057 & 0.048 & 0.055 & 0.057 & 0.055 & 0.059 & 0.032 & 0.026 \\
720 & -0.094 & -0.052 & 0.096 & 0.080 & 0.067 & 0.022 & 0.049 & 0.074 & 0.004 & -0.030 & 0.065 & 0.031 & 0.039 \\
$\sigma$ & 0.149 & 0.151 & 0.091 & 0.094 & 0.076 & 0.058 & 0.044 & 0.041 & 0.041 & 0.032 & 0.036 & 0.019 & 0.014 \\
722 & -0.021 & 0.061 & -0.028 & -0.013 & 0.005 & 0.059 & 0.034 & 0.044 & 0.020 & 0.017 & 0.050 & -0.006 & 0.020 \\
$\sigma$ & 0.048 & 0.063 & 0.051 & 0.046 & 0.037 & 0.028 & 0.024 & 0.026 & 0.027 & 0.022 & 0.026 & 0.014 & 0.011 \\
743 & 0.097 & 0.092 & 0.087 & 0.050 & 0.082 & 0.034 & 0.062 & 0.069 & 0.060 & 0.037 & 0.048 & 0.025 & 0.061 \\
$\sigma$ & 0.015 & 0.020 & 0.013 & 0.013 & 0.011 & 0.009 & 0.008 & 0.009 & 0.009 & 0.007 & 0.008 & 0.004 & 0.004 \\
839 & 0.062 & 0.266 & 0.009 & 0.030 & 0.052 & -0.026 & 0.020 & 0.039 & 0.039 & 0.062 & 0.013 & 0.007 & 0.007 \\
$\sigma$ & 0.064 & 0.092 & 0.080 & 0.060 & 0.051 & 0.044 & 0.038 & 0.043 & 0.044 & 0.044 & 0.051 & 0.027 & 0.021 \\
861 & 0.164 & 0.311 & 0.133 & 0.057 & 0.052 & -0.003 & 0.081 & 0.066 & 0.032 & 0.044 & 0.038 & -0.001 & 0.042 \\
$\sigma$ & 0.078 & 0.107 & 0.082 & 0.063 & 0.050 & 0.040 & 0.032 & 0.035 & 0.038 & 0.037 & 0.041 & 0.022 & 0.017 \\
863 & -0.008 & 0.120 & 0.105 & 0.143 & 0.137 & 0.020 & 0.068 & 0.066 & 0.099 & 0.006 & 0.081 & 0.061 & 0.064 \\
$\sigma$ & 0.152 & 0.193 & 0.114 & 0.115 & 0.090 & 0.069 & 0.061 & 0.066 & 0.066 & 0.059 & 0.066 & 0.035 & 0.027 \\
876 & 0.037 & -0.203 & 0.153 & 0.216 & 0.024 & 0.040 & -0.002 & 0.019 & 0.081 & -0.032 & 0.062 & 0.131 & -0.001 \\
$\sigma$ & 0.168 & 0.262 & 0.161 & 0.173 & 0.130 & 0.100 & 0.096 & 0.086 & 0.087 & 0.076 & 0.093 & 0.047 & 0.034 \\
966 & 0.176 & -0.027 & 0.197 & 0.169 & 0.120 & 0.091 & 0.017 & 0.058 & 0.062 & 0.089 & 0.111 & 0.048 & 0.087 \\
$\sigma$ & 0.146 & 0.222 & 0.132 & 0.135 & 0.096 & 0.082 & 0.074 & 0.075 & 0.076 & 0.065 & 0.071 & 0.038 & 0.032 \\
993 & 0.108 & 0.045 & -0.011 & 0.112 & 0.054 & 0.029 & 0.017 & 0.046 & 0.027 & 0.038 & 0.063 & -0.002 & -0.012 \\
$\sigma$ & 0.107 & 0.147 & 0.112 & 0.100 & 0.086 & 0.066 & 0.058 & 0.066 & 0.067 & 0.061 & 0.068 & 0.036 & 0.030 \\
1029 & 0.097 & 0.090 & 0.071 & 0.056 & 0.065 & 0.033 & 0.062 & 0.067 & 0.052 & 0.036 & 0.042 & 0.026 & 0.054 \\
$\sigma$ & 0.008 & 0.012 & 0.008 & 0.008 & 0.007 & 0.005 & 0.005 & 0.005 & 0.005 & 0.004 & 0.005 & 0.003 & 0.002 \\
1089 & 0.019 & -0.085 & 0.072 & 0.040 & 0.073 & -0.015 & 0.052 & 0.033 & -0.012 & -0.002 & 0.003 & 0.013 & 0.025 \\
$\sigma$ & 0.081 & 0.120 & 0.072 & 0.076 & 0.062 & 0.054 & 0.046 & 0.051 & 0.053 & 0.046 & 0.054 & 0.029 & 0.024 \\
1104 & -0.348 & 0.167 & 0.143 & 0.120 & 0.150 & 0.022 & -0.085 & 0.013 & 0.075 & -0.015 & 0.037 & -0.029 & 0.030 \\
$\sigma$ & 0.358 & 0.418 & 0.221 & 0.198 & 0.166 & 0.121 & 0.113 & 0.090 & 0.096 & 0.079 & 0.092 & 0.048 & 0.034 \\
1154 & -0.019 & -0.113 & -0.018 & 0.032 & 0.122 & 0.023 & -0.006 & -0.050 & 0.115 & 0.047 & 0.071 & 0.042 & -0.011 \\
$\sigma$ & 0.250 & 0.300 & 0.291 & 0.216 & 0.128 & 0.111 & 0.098 & 0.106 & 0.102 & 0.111 & 0.118 & 0.062 & 0.046 \\
1162 & 0.046 & 0.132 & 0.052 & 0.075 & 0.084 & 0.026 & 0.062 & 0.053 & 0.063 & 0.029 & 0.046 & 0.020 & 0.075 \\
$\sigma$ & 0.043 & 0.057 & 0.039 & 0.039 & 0.031 & 0.025 & 0.022 & 0.025 & 0.025 & 0.021 & 0.024 & 0.013 & 0.010 \\
1172 & 0.092 & 0.168 & -0.009 & 0.030 & -0.072 & -0.031 & 0.082 & 0.025 & 0.099 & 0.063 & 0.024 & -0.002 & 0.013 \\
$\sigma$ & 0.138 & 0.212 & 0.150 & 0.134 & 0.116 & 0.088 & 0.071 & 0.085 & 0.083 & 0.076 & 0.090 & 0.045 & 0.035 \\
1257 & -0.067 & 0.185 & -0.107 & 0.123 & 0.041 & 0.101 & -0.001 & -0.002 & 0.044 & -0.065 & -0.145 & 0.062 & -0.032 \\
$\sigma$ & 0.223 & 0.237 & 0.251 & 0.205 & 0.150 & 0.121 & 0.106 & 0.107 & 0.107 & 0.095 & 0.113 & 0.059 & 0.050 \\
1265 & 0.091 & 0.073 & 0.128 & 0.032 & 0.092 & 0.037 & 0.047 & 0.047 & 0.046 & 0.015 & 0.043 & 0.024 & 0.044 \\
$\sigma$ & 0.062 & 0.088 & 0.057 & 0.054 & 0.042 & 0.034 & 0.028 & 0.030 & 0.030 & 0.027 & 0.030 & 0.016 & 0.013 \\
1300 & 0.035 & 0.006 & 0.061 & -0.022 & 0.049 & -0.034 & 0.043 & 0.028 & 0.017 & 0.007 & 0.004 & 0.006 & -0.008 \\
$\sigma$ & 0.104 & 0.136 & 0.090 & 0.083 & 0.076 & 0.057 & 0.041 & 0.044 & 0.044 & 0.034 & 0.039 & 0.020 & 0.015 \\
1301 & 0.207 & 0.100 & 0.140 & 0.163 & 0.064 & 0.030 & 0.092 & 0.075 & 0.077 & 0.017 & 0.042 & 0.038 & 0.081 \\
$\sigma$ & 0.050 & 0.074 & 0.044 & 0.043 & 0.035 & 0.028 & 0.024 & 0.025 & 0.025 & 0.021 & 0.023 & 0.012 & 0.010 \\
1308 & 0.157 & 0.205 & 0.049 & 0.018 & 0.091 & 0.052 & 0.070 & 0.057 & 0.057 & 0.099 & 0.083 & 0.036 & 0.071 \\
$\sigma$ & 0.094 & 0.123 & 0.101 & 0.082 & 0.061 & 0.047 & 0.040 & 0.044 & 0.046 & 0.045 & 0.048 & 0.028 & 0.020 \\
1309 & 0.040 & -0.015 & 0.005 & 0.031 & 0.056 & -0.039 & 0.044 & 0.027 & 0.031 & 0.024 & 0.032 & 0.006 & -0.011 \\
$\sigma$ & 0.033 & 0.047 & 0.035 & 0.036 & 0.030 & 0.026 & 0.023 & 0.026 & 0.026 & 0.022 & 0.026 & 0.014 & 0.012 \\
1327 & 0.089 & 0.152 & 0.043 & 0.055 & -0.054 & -0.010 & 0.082 & 0.053 & 0.044 & 0.048 & 0.062 & 0.007 & 0.005 \\
$\sigma$ & 0.104 & 0.141 & 0.091 & 0.088 & 0.071 & 0.049 & 0.037 & 0.037 & 0.038 & 0.033 & 0.038 & 0.020 & 0.014 \\
1341 & 0.050 & 0.061 & -0.019 & -0.009 & 0.031 & 0.021 & 0.024 & -0.028 & 0.042 & -0.013 & 0.025 & -0.030 & -0.041 \\
$\sigma$ & 0.113 & 0.141 & 0.100 & 0.106 & 0.089 & 0.071 & 0.061 & 0.074 & 0.073 & 0.064 & 0.075 & 0.038 & 0.032 \\
1350 & 0.090 & 0.077 & -0.044 & 0.209 & 0.289 & -0.079 & 0.018 & 0.015 & -0.005 & -0.178 & -0.024 & 0.073 & 0.036 \\
$\sigma$ & 0.217 & 0.298 & 0.268 & 0.239 & 0.164 & 0.161 & 0.148 & 0.168 & 0.162 & 0.150 & 0.153 & 0.083 & 0.073 \\
1352 & -0.016 & 0.041 & 0.045 & 0.019 & 0.046 & 0.009 & 0.018 & 0.024 & 0.003 & 0.013 & 0.009 & -0.002 & 0.007 \\
$\sigma$ & 0.018 & 0.025 & 0.018 & 0.019 & 0.016 & 0.013 & 0.012 & 0.014 & 0.015 & 0.012 & 0.014 & 0.008 & 0.006 \\
1363 & 0.084 & 0.100 & 0.097 & 0.191 & 0.037 & -0.037 & 0.027 & 0.042 & 0.022 & -0.014 & -0.015 & 0.038 & 0.013 \\
$\sigma$ & 0.137 & 0.212 & 0.126 & 0.125 & 0.104 & 0.078 & 0.061 & 0.058 & 0.061 & 0.048 & 0.054 & 0.028 & 0.021 \\
1393 & 0.168 & 0.221 & 0.038 & -0.026 & 0.082 & 0.057 & 0.078 & 0.093 & 0.060 & 0.024 & 0.068 & 0.035 & 0.079 \\
$\sigma$ & 0.074 & 0.114 & 0.088 & 0.059 & 0.044 & 0.036 & 0.030 & 0.032 & 0.035 & 0.036 & 0.040 & 0.022 & 0.016 \\
1413 & 0.094 & 0.163 & 0.130 & 0.027 & 0.114 & 0.005 & 0.044 & 0.034 & 0.115 & 0.018 & 0.056 & 0.033 & 0.068 \\
$\sigma$ & 0.071 & 0.095 & 0.064 & 0.061 & 0.049 & 0.039 & 0.034 & 0.036 & 0.035 & 0.029 & 0.033 & 0.018 & 0.014 \\
1428 & 0.050 & -0.153 & -0.113 & -0.134 & 0.140 & 0.106 & 0.107 & 0.068 & -0.030 & 0.066 & -0.033 & 0.026 & 0.066 \\
$\sigma$ & 0.145 & 0.232 & 0.146 & 0.143 & 0.097 & 0.074 & 0.061 & 0.067 & 0.070 & 0.058 & 0.065 & 0.034 & 0.027 \\
1456 & 0.156 & 0.083 & 0.099 & 0.119 & 0.131 & 0.024 & 0.077 & 0.045 & 0.070 & -0.009 & 0.039 & 0.029 & 0.070 \\
$\sigma$ & 0.094 & 0.118 & 0.094 & 0.068 & 0.058 & 0.048 & 0.040 & 0.045 & 0.046 & 0.043 & 0.048 & 0.026 & 0.020 \\
1490 & -0.020 & 0.089 & 0.041 & 0.042 & 0.048 & 0.011 & 0.027 & 0.012 & 0.043 & -0.013 & -0.010 & -0.018 & 0.025 \\
$\sigma$ & 0.041 & 0.054 & 0.038 & 0.037 & 0.032 & 0.025 & 0.021 & 0.024 & 0.024 & 0.020 & 0.024 & 0.013 & 0.010 \\
1495 & -0.175 & 0.131 & -0.218 & 0.160 & 0.209 & 0.177 & 0.083 & -0.064 & -0.216 & 0.179 & -0.067 & -0.071 & 0.010 \\
$\sigma$ & 0.464 & 0.443 & 0.414 & 0.395 & 0.265 & 0.209 & 0.172 & 0.217 & 0.284 & 0.169 & 0.212 & 0.103 & 0.084 \\
1496 & -0.055 & 0.061 & -0.091 & 0.022 & 0.001 & 0.081 & 0.061 & 0.026 & 0.014 & 0.090 & -0.106 & 0.049 & 0.077 \\
$\sigma$ & 0.140 & 0.172 & 0.125 & 0.107 & 0.089 & 0.071 & 0.064 & 0.063 & 0.061 & 0.050 & 0.059 & 0.030 & 0.024 \\
1506 & -0.242 & 0.332 & -0.011 & 0.076 & 0.030 & 0.065 & 0.110 & 0.037 & 0.033 & -0.021 & -0.005 & 0.208 & 0.128 \\
$\sigma$ & 0.577 & 0.383 & 0.307 & 0.404 & 0.208 & 0.145 & 0.128 & 0.116 & 0.109 & 0.093 & 0.108 & 0.057 & 0.040 \\
1512 & -0.038 & 0.038 & 0.002 & -0.006 & -0.032 & 0.016 & 0.007 & 0.032 & 0.038 & 0.029 & 0.103 & -0.012 & -0.008 \\
$\sigma$ & 0.059 & 0.080 & 0.068 & 0.058 & 0.051 & 0.040 & 0.036 & 0.041 & 0.042 & 0.038 & 0.045 & 0.024 & 0.019 \\
1524 & -0.020 & 0.164 & 0.042 & 0.036 & 0.055 & 0.009 & 0.054 & 0.045 & 0.056 & 0.048 & 0.042 & 0.005 & 0.060 \\
$\sigma$ & 0.050 & 0.065 & 0.045 & 0.045 & 0.035 & 0.028 & 0.023 & 0.026 & 0.026 & 0.021 & 0.025 & 0.013 & 0.011 \\
1537 & -0.035 & -0.051 & 0.067 & 0.053 & 0.041 & 0.027 & 0.038 & 0.011 & 0.001 & 0.030 & 0.036 & 0.033 & 0.002 \\
$\sigma$ & 0.070 & 0.089 & 0.069 & 0.064 & 0.055 & 0.043 & 0.037 & 0.043 & 0.043 & 0.039 & 0.045 & 0.023 & 0.018 \\
1563 & -0.028 & 0.048 & -0.006 & -0.007 & 0.093 & 0.044 & 0.040 & 0.037 & -0.015 & 0.017 & 0.014 & 0.007 & -0.003 \\
$\sigma$ & 0.070 & 0.092 & 0.063 & 0.065 & 0.053 & 0.043 & 0.038 & 0.043 & 0.044 & 0.037 & 0.043 & 0.023 & 0.019 \\
1571 & 0.077 & 0.007 & 0.038 & 0.032 & 0.050 & 0.013 & 0.053 & 0.054 & 0.084 & 0.031 & 0.064 & 0.036 & 0.065 \\
$\sigma$ & 0.045 & 0.061 & 0.042 & 0.041 & 0.031 & 0.024 & 0.021 & 0.022 & 0.023 & 0.020 & 0.023 & 0.012 & 0.010 \\
1627 & 0.042 & 0.008 & 0.034 & 0.033 & 0.041 & 0.022 & 0.043 & 0.055 & 0.038 & 0.016 & 0.025 & 0.016 & 0.047 \\
$\sigma$ & 0.036 & 0.048 & 0.032 & 0.032 & 0.025 & 0.020 & 0.017 & 0.019 & 0.019 & 0.016 & 0.018 & 0.010 & 0.008 \\
1635 & 0.148 & 0.090 & 0.052 & 0.071 & 0.090 & 0.022 & 0.057 & 0.034 & 0.037 & 0.026 & 0.034 & 0.032 & 0.049 \\
$\sigma$ & 0.055 & 0.083 & 0.052 & 0.048 & 0.038 & 0.030 & 0.025 & 0.027 & 0.027 & 0.022 & 0.026 & 0.013 & 0.011 \\
1643 & 0.148 & 0.143 & 0.217 & 0.060 & 0.103 & 0.026 & 0.073 & 0.025 & 0.079 & -0.003 & 0.105 & 0.021 & 0.086 \\
$\sigma$ & 0.110 & 0.134 & 0.086 & 0.081 & 0.061 & 0.049 & 0.041 & 0.041 & 0.042 & 0.037 & 0.041 & 0.022 & 0.017 \\
1652 & 0.018 & -0.053 & -0.037 & -0.011 & -0.005 & 0.070 & 0.014 & -0.087 & -0.021 & -0.032 & -0.046 & 0.006 & -0.005 \\
$\sigma$ & 0.124 & 0.175 & 0.178 & 0.124 & 0.101 & 0.084 & 0.072 & 0.100 & 0.096 & 0.088 & 0.098 & 0.054 & 0.041 \\
1816 & 0.207 & 0.220 & 0.204 & 0.059 & 0.023 & 0.091 & 0.081 & 0.031 & 0.063 & 0.018 & 0.023 & 0.050 & 0.049 \\
$\sigma$ & 0.085 & 0.127 & 0.081 & 0.066 & 0.054 & 0.040 & 0.033 & 0.035 & 0.036 & 0.035 & 0.039 & 0.021 & 0.015 \\
1859 & 0.028 & 0.131 & 0.072 & 0.053 & 0.064 & 0.030 & 0.041 & 0.079 & 0.086 & 0.040 & 0.060 & 0.054 & 0.056 \\
$\sigma$ & 0.059 & 0.079 & 0.054 & 0.049 & 0.037 & 0.029 & 0.025 & 0.027 & 0.028 & 0.029 & 0.032 & 0.017 & 0.013 \\
1946 & 0.126 & -0.178 & 0.056 & 0.122 & 0.039 & 0.044 & 0.059 & 0.116 & 0.014 & -0.004 & -0.018 & 0.038 & 0.081 \\
$\sigma$ & 0.104 & 0.154 & 0.127 & 0.091 & 0.081 & 0.061 & 0.054 & 0.053 & 0.059 & 0.055 & 0.062 & 0.032 & 0.024 \\
1951 & -0.317 & -0.902 & 0.355 & -0.063 & -0.019 & 0.081 & 0.037 & -0.076 & -0.034 & 0.102 & 0.063 & -0.004 & 0.004 \\
$\sigma$ & 0.370 & 0.755 & 0.279 & 0.321 & 0.257 & 0.170 & 0.124 & 0.154 & 0.144 & 0.114 & 0.139 & 0.073 & 0.056 \\
2081 & 0.179 & 0.085 & -0.025 & -0.015 & 0.004 & -0.003 & 0.017 & -0.060 & 0.086 & -0.054 & 0.093 & 0.038 & 0.065 \\
$\sigma$ & 0.139 & 0.168 & 0.170 & 0.115 & 0.093 & 0.076 & 0.064 & 0.074 & 0.074 & 0.076 & 0.084 & 0.047 & 0.034 \\
2087 & -0.085 & -0.062 & -0.048 & 0.029 & 0.036 & -0.040 & 0.033 & 0.017 & -0.098 & -0.037 & 0.022 & 0.015 & -0.039 \\
$\sigma$ & 0.077 & 0.097 & 0.089 & 0.064 & 0.057 & 0.048 & 0.039 & 0.046 & 0.051 & 0.047 & 0.054 & 0.029 & 0.022 \\
2163 & 0.163 & -0.090 & -0.040 & 0.055 & 0.059 & 0.054 & -0.028 & 0.074 & -0.005 & -0.007 & 0.031 & 0.004 & -0.005 \\
$\sigma$ & 0.089 & 0.130 & 0.092 & 0.078 & 0.064 & 0.048 & 0.043 & 0.043 & 0.045 & 0.039 & 0.046 & 0.024 & 0.019 \\
2170 & 0.032 & 0.147 & 0.089 & 0.033 & 0.060 & -0.045 & 0.034 & 0.055 & 0.040 & 0.072 & 0.059 & 0.051 & 0.023 \\
$\sigma$ & 0.065 & 0.091 & 0.062 & 0.055 & 0.044 & 0.035 & 0.030 & 0.033 & 0.034 & 0.034 & 0.038 & 0.020 & 0.016 \\
2171 & 0.000 & -0.068 & 0.025 & -0.008 & 0.100 & 0.009 & 0.088 & 0.054 & 0.055 & 0.007 & 0.055 & 0.033 & 0.028 \\
$\sigma$ & 0.069 & 0.088 & 0.075 & 0.062 & 0.046 & 0.036 & 0.030 & 0.035 & 0.036 & 0.035 & 0.040 & 0.022 & 0.017 \\
2196 & -0.141 & 0.163 & 0.023 & 0.053 & 0.038 & -0.008 & -0.005 & 0.039 & 0.049 & -0.022 & 0.009 & 0.008 & -0.031 \\
$\sigma$ & 0.071 & 0.089 & 0.079 & 0.058 & 0.051 & 0.041 & 0.036 & 0.041 & 0.042 & 0.043 & 0.048 & 0.026 & 0.020 \\
2219 & 0.028 & 0.047 & 0.077 & 0.073 & 0.036 & 0.013 & 0.025 & 0.043 & 0.050 & 0.029 & 0.006 & 0.001 & 0.004 \\
$\sigma$ & 0.048 & 0.069 & 0.054 & 0.046 & 0.039 & 0.032 & 0.027 & 0.031 & 0.032 & 0.028 & 0.032 & 0.017 & 0.014 \\
2230 & 0.035 & 0.039 & 0.054 & 0.025 & 0.034 & -0.002 & 0.037 & 0.059 & 0.043 & 0.020 & 0.026 & 0.017 & 0.013 \\
$\sigma$ & 0.033 & 0.046 & 0.032 & 0.032 & 0.027 & 0.021 & 0.019 & 0.021 & 0.021 & 0.019 & 0.022 & 0.012 & 0.010 \\
2327 & -0.015 & 0.024 & 0.012 & 0.005 & 0.049 & 0.005 & 0.052 & 0.037 & 0.034 & 0.023 & 0.005 & 0.025 & 0.010 \\
$\sigma$ & 0.020 & 0.028 & 0.020 & 0.020 & 0.017 & 0.014 & 0.012 & 0.014 & 0.014 & 0.012 & 0.014 & 0.007 & 0.006 \\
2355 & 0.145 & 0.078 & 0.120 & 0.026 & 0.007 & 0.076 & 0.079 & 0.054 & 0.064 & -0.033 & -0.018 & 0.032 & 0.072 \\
$\sigma$ & 0.098 & 0.128 & 0.110 & 0.078 & 0.064 & 0.050 & 0.043 & 0.046 & 0.048 & 0.051 & 0.057 & 0.031 & 0.023 \\
2381 & 0.163 & 0.171 & 0.123 & 0.102 & 0.073 & 0.014 & 0.083 & 0.084 & 0.077 & 0.020 & 0.020 & 0.043 & 0.066 \\
$\sigma$ & 0.057 & 0.074 & 0.055 & 0.042 & 0.034 & 0.027 & 0.022 & 0.024 & 0.026 & 0.025 & 0.028 & 0.015 & 0.011 \\
2490 & 0.070 & 0.053 & 0.039 & 0.046 & 0.074 & 0.008 & 0.067 & 0.041 & 0.070 & 0.039 & 0.017 & 0.031 & 0.045 \\
$\sigma$ & 0.045 & 0.060 & 0.050 & 0.038 & 0.029 & 0.024 & 0.020 & 0.022 & 0.022 & 0.023 & 0.026 & 0.014 & 0.010 \\
\enddata




\end{deluxetable}

\begin{deluxetable}{ccccccccc}


\tabletypesize{\scriptsize}


\tablecaption{Metallicities of GCs}


\tablehead{\colhead{ID} & \colhead{[Fe/H]} & \colhead{$\sigma_{[Fe/H]}$} & \colhead{ID}  & \colhead{[Fe/H]} & \colhead{$\sigma_{[Fe/H]}$} & \colhead{ID}  & \colhead{[Fe/H]} & \colhead{$\sigma_{[Fe/H]}$}\\ 
\colhead{} & \colhead{(dex)} & \colhead{(dex)} & \colhead{} & \colhead{(dex)} & \colhead{(dex)} & \colhead{} & \colhead{(dex)} & \colhead{(dex)}} 

\startdata
34 & 1.1 & 2.39 & 1327 & -0.95 & 0.76 & 2381 & -0.3 & 0.44 \\
108 & -0.72 & 0.66 & 1341 & -1.49 & 0.52 & 2490 & -0.92 & 0.37 \\
173 & 0.08 & 0.41 & 1350 & -0.88 & 0.5 & PBH30244 & -1.76 & 1.78 \\
187 & -1.51 & 0.36 & 1352 & -1.77 & 0.3 & PBH40083 & -1.29 & 0.8 \\
264 & -1.63 & 0.36 & 1363 & -1.38 & 0.53 & PBH40165 & -1.57 & 0.43 \\
345 & -2.1 & 0.61 & 1393 & -0.12 & 0.27 & PBH40181 & 0.64 & 1.43 \\
359 & -0.43 & 0.83 & 1413 & -0.35 & 0.74 & PBH50037 & -2.34 & 0.83 \\
464 & -0.64 & 0.58 & 1428 & -1.08 & 0.42 & PBH50225 & -0.04 & 0.59 \\
505 & -0.44 & 0.56 & 1456 & -0.65 & 0.62 & PBH50233 & -1.75 & 1.02 \\
526 & -1.04 & 2.02 & 1490 & -1.42 & 0.32 & PBH50286 & -0.04 & 1.85 \\
594 & -0.29 & 0.76 & 1495 & -1.31 & 1.63 & PBH50357 & -3.62 & 2.97 \\
605 & -1.69 & 0.62 & 1496 & -0.4 & 0.93 & PBH50394 & -1.5 & 1.29 \\
628 & -1.17 & 0.52 & 1506 & -0.22 & 1.82 & PBH50401 & -0.04 & 1 \\
676 & 1.15 & 2.96 & 1512 & -1.69 & 0.24 & PBH50696 & -1.86 & 0.5 \\
705 & -0.81 & 0.21 & 1524 & -1.08 & 0.44 & PBH50785 & -0.72 & 1.17 \\
720 & -0.89 & 0.64 & 1537 & -1.85 & 0.36 & PBH50826 & -1.46 & 1.11 \\
722 & -1.44 & 0.35 & 1563 & -1.5 & 0.44 & PBH50886 & -1.79 & 0.87 \\
743 & -0.81 & 0.44 & 1571 & -0.57 & 0.53 & PBH50960 & -1.79 & 0.64 \\
839 & -1.48 & 0.32 & 1627 & -0.9 & 0.36 & PBH51027 & -2.47 & 1.01 \\
861 & -0.57 & 0.46 & 1635 & -0.91 & 0.54 & PBH60045 & -1.03 & 0.97 \\
863 & -0.79 & 0.55 & 1643 & 0.12 & 0.56 & PBH70319 & -2.31 & 1.69 \\
876 & -0.54 & 0.8 & 1652 & -1.81 & 0.44 & PBH70349 & -2.41 & 1.15 \\
966 & -0.37 & 0.35 & 1816 & -0.37 & 0.53 & PBH80172 & -0.77 & 0.68 \\
993 & -0.99 & 0.44 & 1859 & -0.61 & 0.33 & PBH90103 & -2.23 & 0.99 \\
1029 & -0.86 & 0.41 & 1946 & -1.1 & 0.44 & BH91 HS01 & -2.1 & 0.97 \\
1089 & -1.37 & 0.41 & 1951 & -1.94 & 1.11 & BH91 HS02 & -0.92 & 0.4 \\
1104 & -1.26 & 1.44 & 2081 & -1.26 & 0.97 & BH91 HS06 & -1.77 & 0.83 \\
1154 & -0.05 & 0.89 & 2087 & -1.63 & 0.7 & BH91 R012 & 0.19 & 1.44 \\
1162 & -0.82 & 0.45 & 2163 & -1.22 & 0.77 & SBKHP04 & -0.41 & 0.09 \\
1172 & -1.06 & 0.9 & 2170 & -1.08 & 0.42 & SBKHP08 & -0.7 & 0.06 \\
1257 & -1.97 & 0.56 & 2171 & -1.33 & 0.39 & SBKHP09 & -1.21 & 0.13 \\
1265 & -0.8 & 0.43 & 2196 & -1.79 & 0.53 & SBKHP10 & -1.32 & 0.36 \\
1300 & -1.86 & 0.49 & 2219 & -1.64 & 0.43 & SBKHP11 & -1.11 & 0.41 \\
1301 & -0.15 & 0.5 & 2230 & -1.46 & 0.35 & SBKHP12 & -- & -- \\
1308 & -0.4 & 0.53 & 2327 & -1.53 & 0.29 & SBKHP13 & -1.06 & 0.06 \\
1309 & -1.41 & 0.34 & 2355 & -0.26 & 0.51 & SBKHP16 & -0.67 & 0.04 \\
\enddata




\end{deluxetable}

\begin{deluxetable}{ccccccc}


\tablecaption{Linear Fits to $[Fe/H] = a(index)+b$}


\tablehead{\colhead{Index ID} & \colhead{a} & \colhead{b} & \colhead{$R^2$} & \colhead{$R_I$} & \colhead{$\sigma_m$} & \colhead{$\sigma_s$} \\ 
\colhead{} & \colhead{(dex/mag)} & \colhead{(dex)} & \colhead{} & \colhead{} & \colhead{} & \colhead{}} 


\startdata
$\delta$ & 4.12 & -2.68 & 0.884 & 0.2428 & 0.041 & 0.011\\
CNR & 9.52 & -0.96 & 0.845 & 0.1050 & 0.021 & 0.010\\
G4300 & 10.37 & -2.24 & 0.844 & 0.0964 & 0.019 & 0.010\\
MgH & 14.50 & -1.66 & 0.725 & 0.0690 & 0.018 & 0.003\\
Mg2 & 9.83 & -1.89 & 0.886 & 0.1017 & 0.017 & 0.005\\
Mgb & 13.44 & -2.03 & 0.885 & 0.0744 & 0.013 & 0.008\\
Fe5270 & 26.50 & -2.31 & 0.888 & 0.0377 & 0.006 & 0.008\\
Fe5335 & 30.36 & -2.22 & 0.879 & 0.0329 & 0.006 & 0.008\\
Fe5406 & 28.00 & -2.00 & 0.864 & 0.0357 & 0.007 & 0.008\\

\enddata


\end{deluxetable}

\begin{deluxetable}{cccccccc}




\tablecaption{Kinematics of GC Subpopulations}


\tablehead{\colhead{Subpop.} & \colhead{N} & \colhead{V$_c$} & \colhead{V$_r$} &\colhead{V$_{r}$, proj.} & \colhead{PA} & \colhead{$\sigma_V$, uncorr.} & \colhead{$\sigma_V$, rot. corr.} \\ 
\colhead{} & \colhead{} & \colhead{(km s$^{-1}$)} & \colhead{(km s$^{-1}$)} & \colhead{(km s$^{-1}$)} & \colhead{(deg)} & \colhead{(km s$^{-1}$)} & \colhead{(km s$^{-1}$)} } 

\startdata
All      & 108 & -22 $\pm$ 12 & 108 $\pm$  22 & 93 $\pm$ 19 & 261 $\pm$ 11 & 145 & 128 \\
$<$4 kpc & 54 & -19 $\pm$ 21 & 133 $\pm$ 40 & 114 $\pm$ 34 & 294 $\pm$ 14 & 155 & 141 \\
4-8 kpc (incl.) & 37 & -17 $\pm$ 17 & 87 $\pm$ 24 & 75 $\pm$ 21 & 299 $\pm$ 19 & 137 & 133 \\
$>$8 kpc & 17 & -18 $\pm$ 22 & 101 $\pm$ 33 & 87 $\pm$ 28 & 278 $\pm$ 34 & 133 & 132 \\
MR       & 53 & -33 $\pm$ 13 & 122 $\pm$ 18 & 104 $\pm$ 15 & 290 $\pm$ 13 & 148 & 125 \\
MP      & 53* & -36 $\pm$ 19 & 67 $\pm$ 38 & 57  $\pm$ 33 & 323 $\pm$ 21 & 142 & 141 \\
\enddata


\tablecomments{MR GCs are those with [Fe/H] $\geq -1.06$, and MP GCs are those with [Fe/H] $< -1.06$.  MP GC rotation was calculated by excluding a bright outlier, Object 1352, a MP inner GC.  Including this object gave a result inconsistent with the trend outlined by the other MP clusters.}


\end{deluxetable}

\begin{deluxetable}{ccc}




\tablecaption{Projected Mass of M81}


\tablehead{\colhead{N} & \colhead{R$_{max}$} & \colhead{M$_{proj}$} \\ 
\colhead{} & \colhead{(kpc)} & \colhead{($10^{11} M{_\odot}$)} } 

\startdata
54 & 3.82 & $0.88 \pm 0.13$ \\
91 & 7.82 & $1.22 \pm 0.16$ \\
101 & 13.55 & $1.40 \pm 0.22$ \\
108 & 19.67 & $1.64 \pm 0.28$ \\
\enddata




\end{deluxetable}

\end{document}